\documentclass[aps,pre,twocolumn,showpacs,10pt]{revtex4-1}
\usepackage{color}
\usepackage{float}
\usepackage{amsmath}
\usepackage{amsfonts}
\usepackage{amssymb}
\usepackage{graphicx}
\usepackage{float}
\usepackage{textcomp}
\usepackage[dvipsnames]{xcolor}
\usepackage{caption}
\begin{document}

\title{Length scales in Brownian yet non-Gaussian dynamics}
\author{Jos{\'e} M. Miotto$^1$}
\author{Simone Pigolotti$^2$}
\author{Aleksei V. Chechkin$^{3,4}$}
\author{S{\'a}ndalo Rold{\'a}n-Vargas$^{5,6,}$}
\email{sandalo@ugr.es}

\affiliation{$^1$ Leiden Institute of Advanced Computer Science, 2333 CA Leiden, The Netherlands,\\
$^2$ Biological Complexity Unit, Okinawa Institute for Science and Technology and Graduate University, Onna, Okinawa 904-0495, Japan,\\
$^3$ Institute for Physics and Astronomy, University of Potsdam, 14476 Potsdam-Golm, Germany,\\
$^4$ Akhiezer Institute for Theoretical Physics, Kharkov 61108, Ukraine,\\
$^5$ Max Planck Institute for the Physics of Complex Systems, 01187 Dresden, Germany,\\
$^6$ Department of Applied Physics, Faculty of Sciences, University of Granada, 18071 Granada, Spain}

\begin{abstract}

According to the classical theory of Brownian motion, the mean squared displacement of diffusing particles evolves linearly with time whereas the distribution of their displacements is Gaussian. However, recent experiments on mesoscopic particle systems have discovered Brownian yet non-Gaussian regimes where diffusion coexists with an exponential tail in the distribution of displacements. Here we show that, contrary to the present theoretical understanding, the length scale $\lambda$ associated to this exponential distribution does not necessarily scale in a diffusive way. Simulations of Lennard-Jones systems reveal a behavior $\lambda\sim t^{1/3}$ in three dimensions and $\lambda\sim t^{1/2}$ in two dimensions. We propose a scaling theory based on the idea of hopping motion to explain this result. In contrast, simulations of a tetrahedral gelling system, where particles interact by a non-isotropic potential, yield a temperature-dependent scaling of $\lambda$. We interpret this behavior in terms of an intermittent hopping motion. Our findings link the Brownian yet non-Gaussian phenomenon with generic features of glassy dynamics and open new experimental perspectives on the class of molecular and supramolecular systems whose dynamics is ruled by rare events.
\end{abstract}

\maketitle

\section{Introduction}

In one of his celebrated 1905 papers, Einstein proposed a statistical interpretation of Robert Brown's observation based on the corpuscular constitution of matter~\cite{einstein-brownian,robert-brown}. Einstein's theory predicted two concomitant properties for the probability density function (PDF) of displacements of the Brownian particles: its shape must be Gaussian and its variance, the mean squared displacement (MSD), must grow linearly (diffusively) with time. Since the seminal experiments conducted by Perrin more than one hundred years ago~\cite{jean-perrin}, these two predictions were routinely validated and the coexistence between Gaussianity and diffusivity became a paradigm. Exceptions to this long-standing paradigm were first observed in the realm of anomalous diffusion~\cite{random-Bouchaud-Georges,anomalous-Metzler-Klafter,Klages,Sokolov_Softmatter,Hoefling}, where non-linear time dependencies of the MSD coexist with both Gaussian and non-Gaussian PDF's of displacements~\cite{random-Bouchaud-Georges,anomalous-Metzler-Klafter}.\\ 

Recent experiments have found a new class of counterexamples to this paradigm. Several mesoscopic particle systems present a time regime where linear diffusion coexists with a non-Gaussian PDF of displacements 
characterized by an exponential tail $e^{-r/ \lambda (t)}$ as a function of the displacement $r$
~\cite{pnas_granick,nat_mat_granick,hs_granick,MIT_lambda13,Goldstein,Bechhoefer,Metzler_Gaussianity_fair}. The exponential tail is controlled by a time-dependent length scale, $\lambda (t)$, which evolves as a power law: $\lambda (t)\sim t^{\beta}$, with $\beta > 0$. This \textit{Brownian yet non-Gaussian} regime appears in a variety of systems, including colloidal beads moving on the top of lipid tubes~\cite{pnas_granick,nat_mat_granick}, nanospheres in entangled protein suspensions~\cite{pnas_granick}, binary mixtures of colloidal hard spheres~\cite{hs_granick}, microspheres in biological hydrogels~\cite{MIT_lambda13}, and passive tracers in suspensions of eukaryotic swimmers~\cite{Goldstein}.  Some of these works report values for $\beta$ compatible with $1/2$~\cite{pnas_granick,nat_mat_granick,hs_granick,Goldstein} that have motivated theoretical models based on the idea of \textit{diffusing diffusivities}~\cite{Chubynsky_Slater,Jain,Aleksei_PRX,Tyagi,Lanoiselee}.  These models assume a heterogeneous dynamics: particles move according to a time-dependent diffusion coefficient which leads to a Brownian yet non-Gaussian regime with exponential asymptotics~\cite{Chubynsky_Slater,Jain,Aleksei_PRX}. In these models, the variance of the exponential, $\lambda^2 (t)$, is responsible for the total MSD: $\lambda^2 (t) \sim $ MSD$(t) \sim t$. However, more recent experiments have measured a value of $\beta$ significantly smaller than $1/2$~\cite{MIT_lambda13}, challenging an explanation based on diffusing diffusivities. Far from being an exotic phenomenon with an appealing historical background, this problem has deep implications in the understanding of a broad class of systems which are driven by rare events~\cite{nat_mat_granick}.\\ 

Here we study by computer simulations the equilibrium dynamics of representative models of glass-formers with isotropic and non-isotropic interactions~\cite{angell,berthier-biroli-review}. Our main result is that the length scale $\lambda$ associated with the Brownian yet non-Gaussian regime scales in a non-universal way. For systems where particles interact by an isotropic potential, the exponent controlling the evolution of the exponential tail is  $\beta=1/d$, where $d$ is the system dimension. We quantitatively explain this result by a scaling argument based on the idea of hopping motion~\cite{berthier-biroli-review,pinaki_non_gaussian}. In glass-formers where particles interact by a non-isotropic potential, the exponent $\beta$ also depends on the temperature. This dependence becomes more obvious when the system, which behaves as a strong glass-former~\cite{angell}, enters into the Arrhenius regime. We interpret this result as the consequence of an intermittent particle hopping motion at low temperatures. Finally, we show that the dynamics in the Brownian yet non-Gaussian regime for all the explored systems is the result of mixing the anomalous diffusion of individual particles. 

\section{Exponential tail and system dimension}\label{sec:exponentialtail}

We study by Molecular Dynamics simulations the equilibrium dynamics of isotropic Lennard-Jones particles in two~\cite{kob-andersen_2D} and three dimensions by the Kob-Andersen binary mixture~\cite{kob-andersen_original} (see Methods). In both cases, the binary nature of the interaction potential is designed to resulting in a glassy system and, therefore, avoiding crystallization even at very low temperature.  In particular, this system shows the super-Arrhenius behaviour characteristic of fragile glass-formers~\cite{kob-andersen_2D, angell}. This model also shares many fundamental dynamic and structural properties with other representative systems with isotropic interaction. For instance, its relaxation dynamic mechanism (including the emergence of dynamic heterogeneities) and structural patterns are similar to those observed in other molecular liquids~\cite{kob-andersen_original, stillinger-weber}, colloidal hard spheres~\cite{pham_pre, pham_science,berthier_heterogeneities}, models of soft spheres~\cite{soft-spheres}, and even granular materials~\cite{pinaki_non_gaussian}. We cover for this system a wide range of temperatures, from the liquid state (slightly above the system onset temperature) to 1.03 $T_c$, where $T_c$ is the estimated system Mode Coupling Temperature (see Methods).\\

We first measure the particle MSD, see Figure 1a and 1b. For all temperatures, the MSD presents a characteristic short-time local ballistic motion ($\sim t^2$) and a long-time diffusive regime ($\sim t$). Upon cooling, the MSD develops a plateau at intermediate times resulting from an increasingly long local residence time, a common feature of all glass-forming liquids~\cite{berthier-biroli-review}. We characterize the ensemble distribution of displacements by means of the self part of the van Hove function~\cite{hansen}

\begin{equation}
G_s(\vec{r};t) = \frac{1}{N} \left\langle \sum_{i=1}^{N} \delta [\vec{r} - \Delta\vec{r}_i(t)] \right\rangle ,
\end{equation}

\vspace{0.2cm}
\noindent
where $G_s(\vec{r};t)$ is the fraction of particles (from a total number $N$) which have displaced by $\Delta\vec{r}_i(t) = \vec{r}_i(t) - \vec{r}_i(0) = \vec{r}$  in a time $t$. Since both systems are isotropic, we explore the distribution of displacements as a function of the radial coordinate $r=|\vec{r}|$ and define:

\begin{equation}
P (r;t)  \equiv W(r;t)/r^{d-1} = \phi_d G_s(r;t) \,\ ; \,\ d  \in \lbrace 2,3 \rbrace
\end{equation}

\vspace{0.2cm}
\noindent
where $W(r;t)$ is the PDF to have a radial displacement $r$, $d$ the system dimension, and $\phi_d$ the Jacobian angular prefactor ({\em e.g.}, $\phi_2=2\pi$ and $\phi_3=4\pi$)~\cite{hansen}. With this definition we count the fraction of particles which have displaced radially by $r$ normalized (up to prefactors) by $r^{d-1}$ to account for the volume of the shell within the range $r$ to $r+dr$.  In this context, a purely Gaussian diffusion would result in $P (r;t) \sim e^{-r^2/4D(T)t}$, being $D(T)$ the temperature-dependent diffusion coefficient.\\

\begin{figure}[H]
\center
\includegraphics[width=0.95\linewidth]{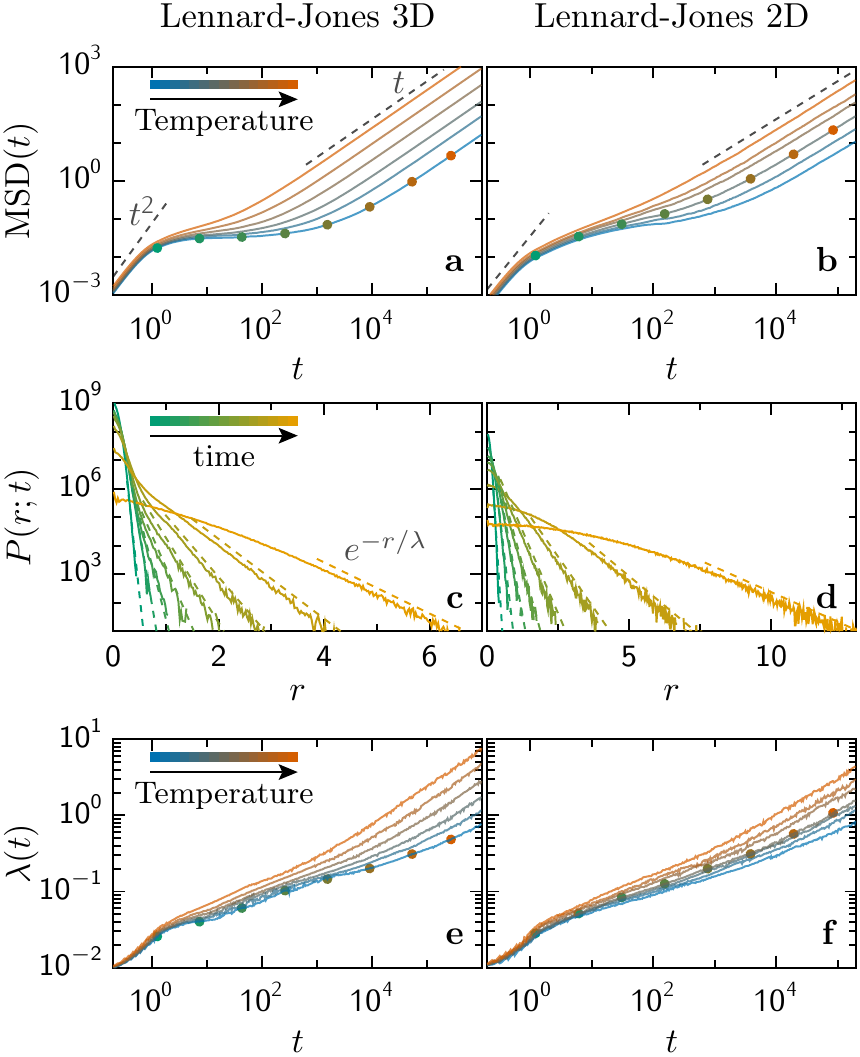}
\caption{\textbf{MSD and exponential tails for the Kob-Andersen binary mixture.}  a) MSD for the 3D system at different temperatures from the liquid state to the deep supercooled regime: $T =$ 0.7, 0.6, 0.54, 0.50, 0.475, and 0.45; b) MSD for the 2D system at different temperatures covering a similar range as in a): $T =$ 0.6, 0.50, 0.45, 0.40, 0.36, and 0.34. The unit of length is equal to the particle diameter (see Methods for the definition of the time and temperature units). Normalized self part of the van Hove function, $P (r;t)$, for the 3D system at $T = 0.45$ (c) and for the 2D system at $T = 0.40$ (d), where the selected times are marked by dots in the corresponding MSD in a) and b). Dashed lines in c) and d) show the extent of the exponential range. Characteristic length $\lambda$ as a function of time for the 3D (e) and 2D (f) systems: in both cases, temperatures are as in a) and b) respectively. Again, dots signal those times corresponding to $P (r;t)$ in c) and d).}
\center
\label{msd_Hove_lambda_LJ}
\end{figure}

The distribution $P (r;t)$ presents a Gaussian behaviour at short distances and an exponential decay at large distances: $P (r;t) \sim e^{-r/ \lambda (t)}$, where $\lambda (t)$ is a characteristic length that increases with time (Fig. 1c and 1d). We observe this exponential decay both at low (3D, Fig. 1c) and intermediate temperatures (2D, Fig. 1d). Qualitatively similar behaviours to those presented in Fig. 1c and 1d appear at all the explored temperatures. This observation is compatible with previous works on glassy systems~\cite{pinaki_non_gaussian}, predicting exponential decays (with logarithmic corrections at large $r$) for the PDF's of displacements.\\

To discriminate the $r$-ranges corresponding to the Gaussian and exponential regimes of $P(r;t)$ and, therefore, accurately estimate $\lambda (t)$, we implemented a non-parametric Kolmogorov-Smirnov test (see Methods). The distinction between the two regimes is more evident at low temperature (in particular at short and intermediate times), being the two regimes separated by an inflection point in $P (r;t)$. At longer times, the Gaussian range extends up to larger distance and therefore becomes dominant. In turn, the range of the exponential tail shrinks and thus becomes marginal at long times. Eventually, this takeover leads to a purely Gaussian distribution of displacements as prescribed by the Central Limit Theorem (CLT).\\

As for the MSD, the growth of $\lambda$  with time is characterized by three distinct regimes (Figs. 1e and 1f). At short times, when the MSD is ballistic, $\lambda (t)$ rapidly increases. Displacements within this short time are typically smaller than the particle diameter (Figs. 1c and 1d) and, therefore, reflect the particle local heterogeneous dynamics. At intermediate times, when the MSD manifests a plateau, $\lambda (t)$ increases in a slower way. In this regime, only a small fraction of particles has abandoned their initial local \textit{cage} and are able to \textit{jump} over distances on the order of few particle diameters (Fig. 1c and 1d). At long times, $\lambda (t)$ increases steeply again. In the same time regime, the MSD is compatible with diffusive behaviour (Fig. 1a and 1b).  In the following, we focus on this Brownian yet-non Gaussian regime~\cite{nat_mat_granick}.\\

Our first goal is to explore the scaling of $\lambda (t)$ within the Brownian yet-non Gaussian regime. We define the time $t_0$ at which the 2D and 3D Lennard-Jones systems start to be diffusive (MSD($t \geq t_0$) $\sim t$). In practice, we define $t_0$  as the time at which the exponent $\mu$ characterizing the scaling of the MSD with time is equal to 1 up to a tolerance of $5\%$, see Methods for details. We compare the scaling with time of MSD$(t)^{1/2}$ and $\lambda(t)$ in the time regime $t \geq t_0$ for different temperatures (Fig. 2). In both systems, the scaling of the MSD is diffusive, MSD$^{1/2} (t \geq t_0) \sim t^{1/2}$. In contrast, the scaling of $\lambda(t)$ is dimension dependent:

\begin{equation}
\lambda (t \geq t_0) \sim t^{\beta(d)} 
\end{equation}

\vspace{0.1cm}
The  exponent $\beta$ does not appreciably depend on the temperature within the diffusive time window $t \geq t_0$. However, $\beta$ clearly depends on the system dimension, with a value compatible with $\beta=1/2$ in two dimensions and $\beta=1/3$ in three dimensions. While the value $\beta=1/2$ in two dimensions is consistent with the diffusive scaling of the  MSD, the scaling of $\lambda$ in three dimensions does not seem to be trivially related with that of the MSD. The fact that the scaling of $\lambda$ and the MSD are not in general trivially linked descends from the fact that the weight of the exponential tail compared to the rest of the distribution is time-dependent. While $\lambda$ is a property of the exponential tail only, the MSD is determined by the entire distribution. In this respect, the relative contribution of the exponential tail to the MSD diminishes, with the start of the exponential range moving to larger and larger displacements, as time increases. The scaling of $\lambda$ also contrasts with the interpretation of the Brownian yet non-Gaussian regime  based on diffusing diffusivities~\cite{nat_mat_granick,Chubynsky_Slater,Aleksei_PRX}, where $\lambda^{2} (t) \sim$ MSD $(t)$ $\sim t$. However, the time evolution we observe for $\lambda (t \geq t_0)$ in three dimensions agrees with recent experiments on microspheres diffusing in biological gels~\cite{MIT_lambda13}, where $\beta < 1/2$ and the total PDF of displacements presents a Gaussian core, which grows with time, and a marginal exponential tail, which tends to disappear with time.\\

\begin{figure}[H]
\center
\includegraphics[width=0.95\linewidth]{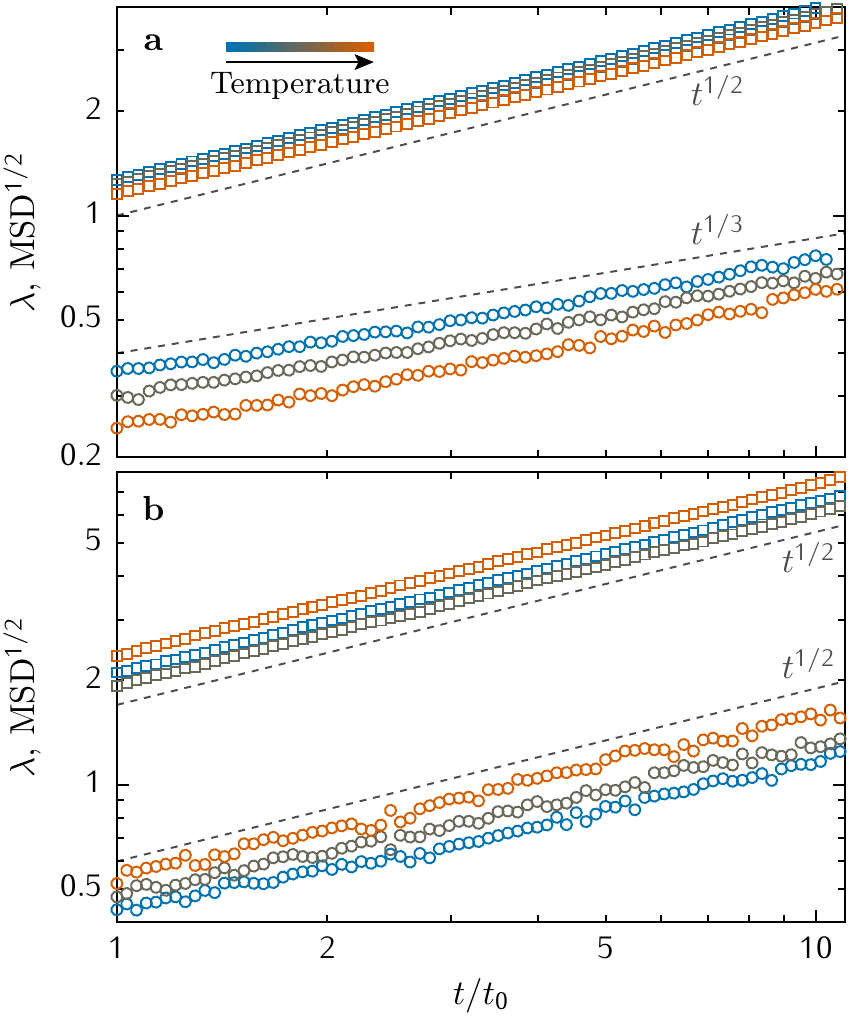}
\caption{\textbf{Brownian yet non-Gaussian regime for the Kob-Andersen binary mixture.} Double log plot of MSD$^{1/2}$ (squares) and $\lambda$ (circles) as a function of time for the 3D (a) and 2D (b) Kob-Andersen systems at different temperatures: $T =$ 0.45, 0.5, and 0.6 (3D), and $T =$ 0.4, 0.5, and 0.6 (2D). Here $t_0$ is the temperature- and system- dependent time at which each system reaches a $\mu$ value compatible with 1 within numerical uncertainty, \textit{i.e.} MSD($t \geq t_0$) $\sim t^{\mu \, \cong \,  1}$. To determine $t_0$ we numerically compute the local exponent $\mu (t)$ controlling the time evolution of the MSD$(t)$, \textit{i.e.} MSD$(t) \sim t^{\mu (t)}$  (see Methods). Dashed lines in a) and b) serve as a reference.}
\center
\label{Fig_msd_lambda_LJ}
\end{figure}

\vspace{1cm}
These observations suggests a general dependence $\beta(d)=1/d$, with $d  \in \lbrace 2,3 \rbrace$. We propose a theory for such scaling behaviour as an outcome of hopping motion, a signature of all glass-forming liquids below the onset temperature~\cite{berthier-biroli-review}. The exponential tail of $P (r;t)$ originates from a minority of particles able to escape from their initial cage and perform a large displacement for times of the order of $t_0$. We call these particles \textit{hoppers} and denote by $N_h(t)$ their number at time $t$. Since the system is at equilibrium, we assume that hoppers escape from their cage at a constant rate $\omega$:

\begin{equation}\label{eq:omega}
N_h(t) \approx \omega N t \sim t .
\end{equation}

\vspace{0.3cm}
\noindent
Equation \eqref{eq:omega} is valid in a time range where $N_h(t)\ll N$, being, therefore, $\omega^{-1} \gg t$. We now use that the functional form of the mass density of hoppers, $\rho_h (r;t)$, as a function of the distance at time $t$ coincides with that of $P (r;t)$ for $t\approx t_0$ and is, therefore, exponential:
\vspace{0.3cm}
\begin{equation}
P (r;t) \sim \rho_h (r;t) = \rho_0 e^{-r/ \lambda (t)} \,\,\,\ ;  \,\,\,\ t \approx t_0
\end{equation}

\vspace{0.3cm}
We consider $\rho_0$ to be independent of time for times of the order of $t_0$, \textit{i.e.} the source of hoppers at $r=0$ remains at constant density for $t \approx t_0$ since $N_h(t)\ll N$. This is consistent with the behaviour of $P (r;t)$ observed in Fig. 1c and 1d, and also clearly manifested by the $P (r;t)$ shown in Appendix C, where all the exponential tails at different times cross at a common value at $r \approx 0$. The number of hoppers at a time $t \approx t_0 $ therefore scales as
\begin{equation}
\begin{split}
& N_h(t) \sim \int_{0}^{\infty}  r^{d-1} \rho_h (r;t)  \,\ dr  \sim \\ 
& \,\,\,\,\,\,\,\,\,\,\,\,\ \sim \int_{0}^{\infty} r^{d-1} e^{-r/ \lambda (t)} \,\ dr \sim \lambda^d(t)
\end{split} 
\end{equation}

\vspace{0.5cm}
Combining Eq.~\eqref{eq:omega} and Eq. (6) we finally obtain
\begin{equation}
\lambda (t) \sim t^{1/d}
\end{equation}

\vspace{0.3cm}
The scaling argument embodied in Eq (4) to (6), and leading to Eq. (7) would, in principle, hold for all $d$. In particular, we have also tested Eq. (7) in Appendix A for $d=4$ for the Lennard-Jones system investigated, also confirming its validity. Nevertheless, the general extension of our scaling argument needs subsequent investigation. In particular, works based on a Mode Coupling Theory formalism predict changes in the dynamics of four-point correlators at higher dimensions whose hypothetical effect on the observables studied here should be investigated~\cite{Biroli-Bouchaud}. Apart from that, and due to the observed emergence of finite size effects in 2D simulations~\cite{size-2D-1,size-2D-2,size-2D-3}, we have tested and confirmed Eq. (7) for a much more larger 2D-Lennard-Jones system in Appendix B, in accord with recent experiments in 2D systems~\cite{pastore-prl-2021}.\\ 

Summarizing, our argument to explain $\lambda (t)\sim t^{1/d}$ is based on three assumptions: (1) hoppers leave their local environment at a fixed hopping rate (Eq. (4)); (2) hoppers fill the space isotropically (Eq. (6)); (3) the restriction to a time regime $t \approx t_0$ where $N_h(t) \ll N$. We test these assumptions individually for $d=2,3$ in Appendix C. Once the majority of the particles have abandoned their initial position ($t \gg t_0$), and present a statistically equivalent jumping record, the last assumption breaks. At such a long time, the PDF of displacements results from a large number of independent and equally distributed displacements and is, therefore, Gaussian by virtue of the CLT.

\section{Exponential tail and non-isotropic interactions}

We now investigate the evolution of $\lambda (t)$ when particles interact by a non-isotropic potential. To this purpose we consider a three-dimensional system of tetravalent \textit{patchy particles}~\cite{rovigatti_molphys,lorenzo-patchy,dnapatchy1}. Such particles have four sticky spots tetrahedrally distributed on their surface which provide a strongly directional interaction with fixed valence. Upon cooling, the dynamics slows down and the system develops an amorphous, highly connected tetrahedral network~\cite{soft_matter_tetrahedral}. This model has already shown its capability for capturing some fundamental dynamic and structural features, such as the emergence of an amorphous tetrahedral order, present in different classical systems with non-isotropic interactions such as atomistic models of water, silicon, and silica~\cite{rovigatti_molphys, soft_matter_tetrahedral, water1_ff, amorphous-ice-sq, silicon-sq, horbach_walter_silica}. We investigate the equilibrium dynamics of this system, which shows the Arrhenius behaviour characteristic of a strong glass-former~\cite{soft_matter_tetrahedral, angell}, at moderate density by Brownian Dynamics simulations (see Methods) and explore the emergence of the Brownian yet non-Gaussian regime. Our study covers a wide $T$-range from the liquid state (slightly above the percolation threshold) to the deep Arrhenius regime, where the great majority of the particles are tightly bound to their neighbors~\cite{rovigatti_molphys}.\\

In contrast to the Lennard-Jones systems, the scaling of $\lambda(t)$ clearly depends on $T$ for the tetrahedral gelling system (Figure 3). While at high temperature $\lambda(t)  \sim t^{1/3}$, compatible with the 3D Lennard-Jones system, the exponent $\beta$ decreases upon cooling the system below the Arrhenius temperature~\cite{rovigatti_molphys,soft_matter_tetrahedral}, reaching values $\beta\approx 1/6$ for temperatures deeply into the Arrhenius regime. Thus our argument leading to $\beta(d) = 1/d$ independently of $T$ (Eq. (4) to (7)) does not hold for the gelling system at low $T$. To understand this discrepancy, we individually test the hypotheses underlying our scaling theory (see Appendix C). We find that in this system Eq. (4) is not satisfied for the gelling system: in particular, the number of hoppers appears to grow sublinearly with time. This observation points to a scenario where the production of hoppers becomes more and more intermittent as the temperature decreases. Such phenomenon seems to be absent in the Lennard-Jones systems, where our scaling theory holds in all the range of temperatures we explored (see Appendix). We stress that Eq. 6 is satisfied by our tetrahedral gelling system (see Appendix C), despite the system could display a fractal structure~\cite{soft_matter_tetrahedral}. This result should however be confirmed in other systems showing non trivial structures ~\cite{Meyer, Spanner}.  \\

\begin{figure}[H]
\center
\includegraphics[width=0.95\linewidth]{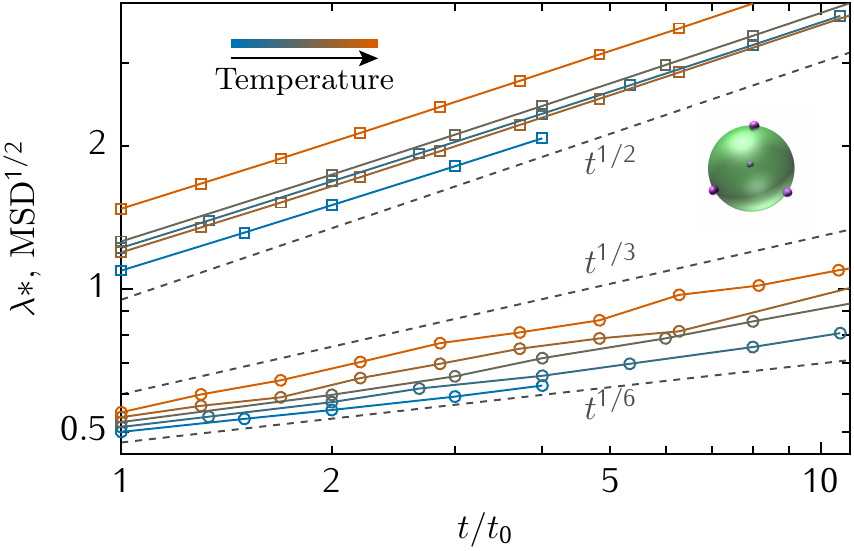}
\caption{\textbf{Brownian yet non-Gaussian regime in the tetrahedral patchy system.} Double log plot of MSD$^{1/2}$ (squares) and $\lambda_*$ (circles) as a function of time for the tetrahedral patchy system at different temperatures. To make evident the spread of curves at different temperatures, we present a rescaled value $\lambda_*$ for the different temperatures to make them start from an almost common value at $t_0$. As in Fig.2, $t_0$ is the temperature-dependent time at which the tetrahedral system reaches a  value of $\mu$ compatible with 1. We cover a wide $T$-range from the liquid state to the deep Arrhenius regime: $T =$ 0.1025, 0.105, 0.11, 0.115, and 0.14 (see Methods). The Arrhenius temperature, \textit{i.e.} the temperature at which the $T$-dependence of the diffusion coefficient becomes exponential, is  $T_A\approx 0.115$~\cite{rovigatti_molphys,soft_matter_tetrahedral} (see also Methods and Figure 4). Dashed lines serve as a reference. The unit of length is taken as the patchy particle hard-sphere diameter (see Methods). The figure also shows a sketch of a tetrahedral patchy particle.} 
\center
\label{Fig_msd_lambda_gel}
\end{figure}

We summarize the behaviour of the exponent $\beta(T;d)$ for the three systems  (see Figure 4).  For the two Lennard-Jones systems, $\beta$ markedly depends on the system dimension (Eq. (3)) and is practically independent of $T$ (with, at most, a very modest decrease in 3D). Our theoretical argument (Eqs. (4) to (7)) predicts that other systems with isotropic interactions, such as hard and soft spheres, should present the same behaviour. Instead, $\beta$ clearly depends on temperature for the tetrahedral gelling system. This dependence is stronger once the system enters into the Arrhenius regime, where the diffusion coefficient decreases exponentially upon cooling~\cite{soft_matter_tetrahedral} (inset in Fig.4), thereby revealing a connection between a defining feature of the dynamics observed in strong glass-formers and the emergence of rare events. At higher temperatures, where connectivity is low~\cite{rovigatti_molphys}, $\beta$ attains a value compatible with that of the 3D Lennard-Jones system ($\cong 1/3$).\\

\begin{figure}[H]
\center
\includegraphics[width=0.96\linewidth]{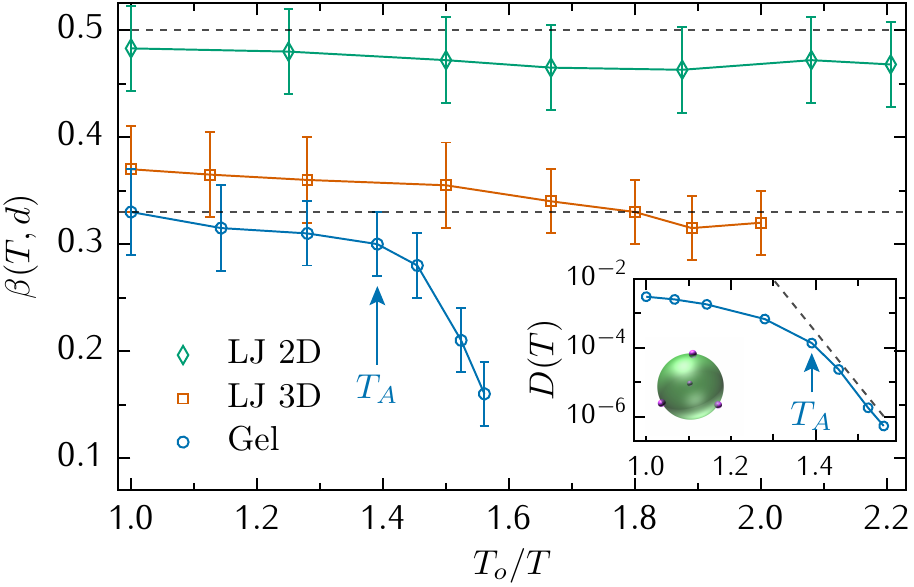}
\caption{\textbf{Compendium of $\beta$ exponents.} $\beta(T;d)$ in the time regime $t \geq t_0$ as a function of the temperature for the three systems investigated. Here $T_o$ stands for the onset temperature for the two Lennard-Jones systems and for the percolation temperature for the tetrahedral gelling system (see Methods). The explored $T$-range goes from the liquid state to the deep supercooled regime. Horizontal dashed lines serve as a reference for $1/2$ and $1/3$. A vertical arrow signals the temperature $T_A$ at which the tetrahedral gelling system enters into the Arrhenius regime~\cite{soft_matter_tetrahedral}. Inset: Diffusion coefficient $D(T)$ (obtained from the measured MSD) as a function of the temperature for the tetrahedral gelling system (dashed line serves as a reference for the exponential, low-$T$, Arrhenius behaviour which starts at $T_A$)~\cite{soft_matter_tetrahedral}.}
\center
\label{Fig_msd_lambda_gel}
\end{figure}

\section{Non-Gaussianity and dynamics by populations}

We now look further into the coexistence between linear diffusion and non-Gaussian distributions of displacements in 3D. To this aim, we discriminate the particles into populations according to their potential energy $E$ at $t=0$ as in Ref. \cite{soft_matter_tetrahedral}. For the tetrahedral gelling system, this procedure leads to five different particle populations characterized by the number of bonds per particle ($i \in \lbrace 0,1,2,3,4 \rbrace$) at $t=0$, being $E|i$ the potential energy of population $i$ (where $E|i > E|j$ when $i<j$)~\cite{soft_matter_tetrahedral}. For the 3D Lennard-Jones system we arbitrarily define two populations with a marked difference in their initial potential energies: one including the $1\%$ of particles with the highest potential energy and the other one including the $1\%$ of particles with the lowest potential energy, both at $t=0$.\\

The time evolution of the MSD for the different populations at fixed $T$ is presented in Figures 5a and 5b. Populations with high potential energy at $t=0$ show a super-diffusive regime at short times which is compensated with the sub-diffusive motion associated to the populations with low potential energy. In particular, for the tetrahedral system we see that this sub-diffusive motion is dominant and represents almost the total MSD. The reason is that, for the chosen temperature, the system is already highly percolated~\cite{soft_matter_tetrahedral} and the populations having a low energy at $t=0$ are much more larger (number of particles) than the populations having a high energy at $t=0$. Thus, while particles with high potential energy are briefly connected to the gel network, low energy particles are retained for a longer time to the network, resulting in a plausible source for sub-diffusion. At intermediate times, the distinct populations reach different values of the MSD at their respective plateaus. This difference is more pronounced for the tetrahedral liquid due to its lower density. When abandoning the plateau, populations with high potential energy at $t=0$ show sub-diffusive motion while populations with low potential energy at $t=0$ present super-diffusive motion. This is particularly clear when looking at the $\mu(t)$ exponent obtained from the slope of the MSD: when leaving the plateau,  populations starting with a high potential energy present $\mu(t)<$ 1 whereas populations starting with a low potential energy show $\mu(t)>$ 1 (Figures 5c and 5d).\\

The mixing of anomalous diffusions is concomitant with a change in the energy of each population: particles starting with a high (low) potential energy show a decrease (increase) in their energy which slows down (boosts) their dynamics (Figures 5e and 5f). At long times, once the particles lose the memory of their initial state and pass through all the possible energy states, all the populations show the same average energy while their dynamics converge to a common diffusive trend. It is worth highlighting that this convergence happens at very long times pointing out to the existence of a long term memory of the potential energy. In addition, this convergence shows a common {\em relaxation time} which is independent of the particle population. All this phenomenology reduces to a simple intuition: when a particle is in a high (low) energy state it moves faster (slower) than the average. For large times, when all the particles have passed through all the possible fast and slow states, the sampled distributions of displacements for all populations are equivalent and the total PDF of displacements becomes Gaussian as dictated by the CLT.\\

Before collapsing into a common diffusive trend, there exists a time window where the different populations still show $\mu(t)|i \neq 1$ $\forall i$, with $\mu(t) \cong 1$ for the total number of particles (shaded region in Fig. 5). This is the Brownian yet non-Gaussian window where the exponential tail is still detectable despite the system as a whole shows $\mu(t) \cong 1$. This observation helps understanding why within this \textit{diffusive} time window, $P (r;t)$ is not necessarily Gaussian: the distinct populations have not yet converged and, therefore, their sampled dynamic states are not yet equivalent, resulting in a non-Gaussian PDF of displacements. Our results show that the observed exponential tail originates from \textit{mixing anomalies}: within the Brownian yet non-Gaussian regime each individual population shows its own anomalous diffusion ($\mu(t)|i \neq 1$). We have further shown that, within the Brownian yet non-Gaussian regime, the anomalous diffusion of each population (which is a dynamic feature) can be associated to its initial potential energy (which is a structural feature).

\begin{figure}[H]
\center
\includegraphics[width=0.95\linewidth]{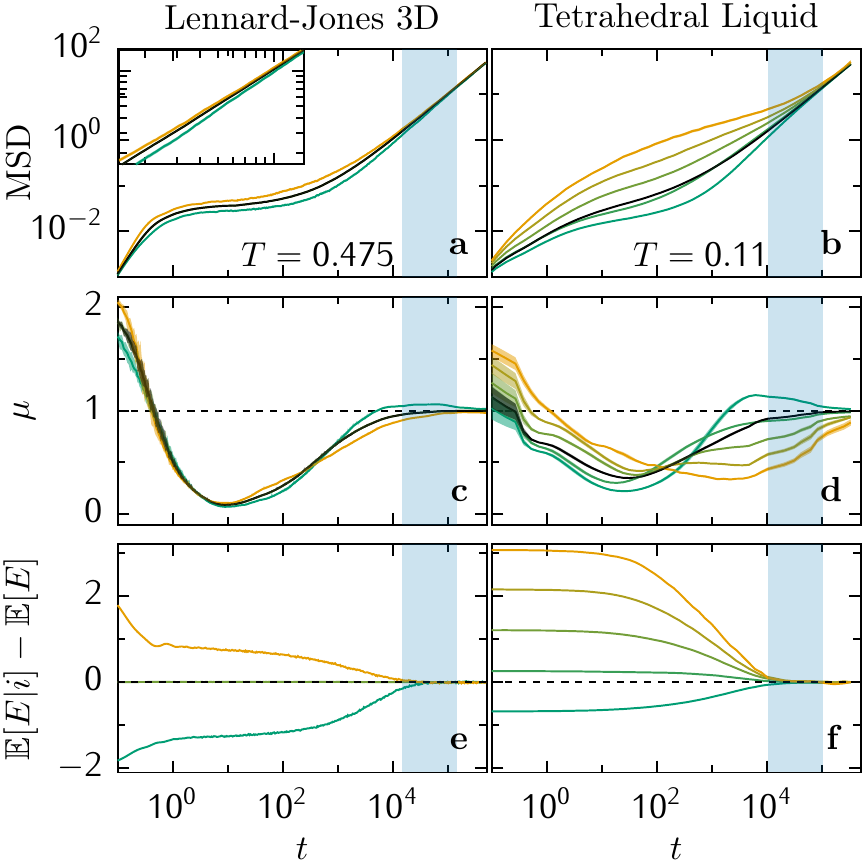}
\caption{\textbf{Dynamics by particle populations.} Left column corresponds to the 3D Kob-Andersen binary mixture and shows the time evolution of different observables discriminated into two particle populations: a) MSD$(t)$, c) $\mu(t)$ (MSD$(t) \sim t^{\mu (t)}$), and (e) difference between the average potential energy of each population, $\mathbb{E}[E|i]$, and the system average potential energy, $\mathbb{E}[E]$. Yellow (green) lines correspond to the population that includes the $1\%$ of the particles with the highest (lowest) potential energy at $t=0$. Right column corresponds to the tetrahedral gelling system and shows the same observables as those appearing in the left column (b, d, and f). The five colored lines correspond to the five populations having a different number of bonds ($i \in \lbrace 0,1,2,3,4 \rbrace$) at $t=0$  (lines from yellow to green). Black lines in the whole figure represent observables  averaged over all particles. The shaded regions mark out the extent of the Brownian yet non-Gaussian time window, where the exponential tail is still detectable and $\mu(t) \cong 1$ for the total number of particles. Inset in a) shows a detail of the MSD corresponding to the shaded time window. The chosen temperatures correspond to the deep supercooled regime (3D Kob-Andersen binary mixture) and the Arrhenius regime (tetrahedral liquid). For both systems, the different populations evolve in such a way that the corresponding observables pass from a short-time behaviour (where the particles still retain the information of their initial mechanical state) to a long-time regime, where the particles have lost memory of their initial state and the different populations converge to a common trend.}
\center
\label{Fig_msd_lambda_gel}
\end{figure}

\section{Conclusions and Perspectives}

Brownian yet non-Gaussian transport has recently attracted great interest for its unexpected but ubiquitous presence. It is at the heart of the more general problem of understanding, in a comprehensive way, the rare event dynamics present in many complex systems. Our study brings new insights into this appealing phenomenon. First, we have shown numerically that for canonical liquid models where particles interact by an isotropic potential, the exponent controlling the evolution of the observed exponential tail is not universal but depends on the system dimension: $\beta (d) = 1/d$. This important finding contrasts with the universality predicted by the current theoretical interpretations. We have rationalized this observation by a theoretical scaling argument which does not depend on the specific functional behaviour of the interaction potential. For that reason we expect other representative systems with significantly different potentials, such as hard and soft spheres, to show a similar scaling behaviour. Second, we have shown that dimension is not the only factor affecting the evolution of the exponential tail. In systems where particles interact by a non-isotropic potential, $\beta$ also depends on temperature. Contrary to the isotropic models investigated in this work (which behave as fragile glass-formers), the specific non-isotropic system investigated here behaves as a strong glass-former. Therefore, the temperature dependence shown by $\beta$ for the non-isotropic system reveals a new fundamental feature discriminating the dynamics of fragile  and strong glass-formers, with far-reaching consequences for understanding and classifying glasses. It also establishes a connection between the emergence of the generic Arrhenius regime representative of strong glass-formers and their hopping dynamic mechanism. Third, we have shown that the time regime where the Brownian yet non-Gaussian transport occurs is characterized by a mixed anomalous diffusion of different particle populations. These populations are characterized by a configurational property: their potential energy. We expect our findings for the models investigated here to hold for a large variety of systems. In this respect, the canonical systems we have chosen in our study share a common fundamental structural and dynamic phenomenology with other representative systems defined by different isotropic and non-isotropic potentials, \textit{e.g.} hard or soft spheres~\cite{pham_pre, pham_science, soft-spheres}, and water, silicon, or silica~\cite{water1_ff, amorphous-ice-sq, silicon-sq, horbach_walter_silica}.\\

Taken in a broad sense, our results bring new research perspectives, imposing severe constraints to future theories and calling for new experiments. Future theoretical models should account for the dimension and temperature dependencies that we observed in this work. For instance, the seminal work by Chaudhury and coworkers~\cite{pinaki_non_gaussian} and the large deviation model proposed by Barkai and Burov~\cite{theory-non-diff-diif-3}, predict exponential tails in the PDF of displacements (with logarithmic corrections at large $r$) using a Continuous Time Random Walk (CTRW) formalism. These two approaches do not suggest a dimension-dependent behavior of the characteristic length $\lambda \sim t^{1/d}$. Other works employing the idea of diffusing diffusivity~\cite{Chubynsky_Slater,Jain,Aleksei_PRX,Tyagi,Lanoiselee,Sposini} lack a dependence on dimension as well. Similar conclusions hold for other approaches to Brownian yet non-Gaussian diffusion, {\em e.g.} models mixing CTRW and diffusing diffusivity~\cite{Song}, polymerization models where the fluctuating size of a diffuser generates a diffusing diffusivity process~\cite{Baldovin,Hidalgo-Soria}, diffusion in a fluctuating corrugated channel~\cite{Li}, and collosal Brownian yet non-Gaussian diffusion induced by nonequilibrium noise~\cite{Bialas}. Future models should also explain the mixed anomalous diffusion observed in our numerical simulations during the Brownian yet non-Gaussian time regime. In our view, this is a crucial yet challenging task. We speculate on the idea of incorporating a CTRW formalism with correlated jumps, one feature that is usually missing in previous works.\\

From a practical perspective, our results can \textit{a priori} be tested in several real systems regulated by different control parameters, not necessarily temperature. For instance, we expect colloidal glass- and gel-forming liquids~\cite{berthier-biroli-review,pinaki_non_gaussian} to present a similar phenomenology as the one described in this paper when studied as a function of other parameters, \textit{e.g.} packing fraction. In addition, non-Gaussian tails have also been reported recently in granular systems~\cite{Walter_granular}, where shear stress plays the role that temperature plays in the systems we studied here. Complex biological media, controlled by pH, have also shown Brownian yet non-Gaussian transport, where the measured value of $\beta$ is compatible with our results~\cite{MIT_lambda13}. Finally, we expect a similar phenomenology as the one reported in this work in non-equilibrium active systems, where transport is controlled by an energy of metabolic origin~\cite{Ramaswamy-review-active,Marchetti-review-active}. In this respect, non-Gaussian regimes have indeed been found in active biological systems constituted by animate, energy-consuming, \textit{particles}, for example in cell migration processes~\cite{cell_migration-pnas} and in the transport of different organelles in the cytoplasm of animal cells~\cite{cytoplasm-cell}.

\section{Methods}

\subsection{Lennard-Jones system} 

We performed two and three dimensional Molecular Dynamics simulations of a Kob-Andersen binary mixture ~\cite{kob-andersen_2D}. In both cases we first run simulations in the canonical ensemble to equilibrate the system at a fixed temperature. From them, we run simulations in the microcanonical ensemble to evaluate the dynamic observables presented in this article. The interaction between a particle of species $\alpha$ and a particle of species $\beta$  is given by the Lennard-Jones potential:\\

\begin{equation}
V_{\alpha\beta}(r) = 4\epsilon_{\alpha\beta} \Big[ \Big( \frac{\sigma_{\alpha\beta}}{r} \Big)^{12}-\Big( \frac{\sigma_{\alpha\beta}}{r} \Big)^{6} \Big]  \,\ ; \,\  \alpha,\beta \in \lbrace A,B \rbrace  ,
\end{equation}
\\

\noindent
where $A$ and $B$ are the labels for the two species and $r$ is the distance between the centers of mass of the two particles. For both systems, $\sigma_{AA} = 1$, $\sigma_{AB} = 0.8$, $\sigma_{BB} = 0.88$, $\epsilon_{AA} = 1$, $\epsilon_{AB} = 1.5$, and $\epsilon_{BB} = 0.5$. The potential is truncated and shifted at $r = 2.5 \sigma_{\alpha\beta}$~\cite{kob-andersen_original}. All the results are given in reduced units, where $\sigma_{AA}$ is the unit of length, $\epsilon_{AA}$ the unit of energy, and $\sigma_{AA}\sqrt{m/48\epsilon_{AA}}$ the unit of time (being $m=1$ the mass of the particles). Temperature, $T$, is controlled during the equilibration process by an Andersen thermostat with an effective mass of 48 reduced units~\cite{kob-andersen_2D} with Boltzmann's constant set to $1$.\\

The number of particles for each species of the 2D system are $N_A^{2D} = 6500$ and $N_B^{2D} = 3500$ with a total number density $\rho^{2D} = (N_A^{2D} + N_B^{2D})/L^2 = 1.16$, being $L = 92.78$ the length of the square simulation box~\cite{kob-andersen_2D}. For the 3D system we used a different composition with $N_A^{3D} = 6400$ and $N_B^{3D} = 1600$ for a total number density $\rho^{3D} = (N_A^{3D} + N_B^{3D})/L^3 = 1.20$, being $L = 18.80$ the length of the cubic box~\cite{kob-andersen_original}. These compositions avoid the emergence of a crystal structure even at very low temperature. We covered temperature ranges $T \in [0.34, 0.75]$ (2D) and $T \in [0.45, 0.9]$ (3D). For both systems (2D and 3D), the lowest temperature investigated is 1.03 $T_c$, where $T_c$ is the Mode Coupling Temperature of the glass transition as estimated from data coming from numerical simulations~\cite{kob-andersen_2D,kob-andersen_PRE-1}. The onset temperatures for the 2D and 3D systems are: $T_o^{2D} = 0.75$~\cite{Flenner-Szamel-2D} and  $T_o^{3D} = 0.9$~\cite{Coslovich-Pastore-onset} (Fig. 4 in the main text). Both systems show at low $T$ the super-Arrhenius dynamic behaviour characteristic of fragile glass-formers~\cite{kob-andersen_2D}. We used for both systems a Velocity Verlet algorithm with a time step depending on the temperature: $\delta t^{2D}=0.02$ ($\delta t^{2D}=0.01$) for $T \leq 0.6$ ($T > 0.6$) and $\delta t^{3D}=0.02$ ($\delta t^{3D}=0.01$) for $T \leq 0.7$ ($T > 0.7$). The runs extended over $10^7$ and $5 \cdot 10^7$ time steps for the 2D and 3D system respectively.\\ 

The results shown in this article correspond in both cases to particles of the species $A$. For the 2D system, we averaged over $50$ independent simulations (around 325.000 individual particle trajectories) for each value of the temperature running on a high-end CPU processor cluster with a total amount of CPU time of 11.5 years.  For the 3D system, we averaged over $70$ independent simulations (around 450.000 individual particle trajectories) for each value of the temperature with a total amount of CPU time of 18.5 years.   

\subsection{Tetrahedral system} 

We performed three dimensional Brownian Dynamics simulations of tetravalent patchy particles in the canonical ensemble. The number of particles is $N=10000$ with a length of the cubic simulation box $L=25.98\;\sigma$, being $\sigma$ the particle hard sphere-like diameter, here taken as the unit of length. The number density is $\rho=N/L^3=0.57$. For this density, the system develops a homogeneous amorphous tetrahedral network even at very low temperature~\cite{fully_bonded_tetra,soft_matter_tetrahedral}. The interaction potential comprises a spherical steep repulsion and a short-range attraction. The interaction between a generic pair of particles $1$ and $2$ is given by:

\begin{equation}
V(1,2) = V_{CM}(1,2) + V_P(1,2)
\end{equation}
\\
where $V_{CM}(1,2)$ is the repulsive part of the potential between particles $1$ and $2$ whereas $V_P(1,2)$ is the attractive part of the potential between the patches of particles $1$ and $2$. These potentials are modeled as follows:

\begin{equation} 
V_{CM}(12) = {\left( \frac{\sigma}{r_{12}}\right)}^{p} \,\,\,\,\,\,\,\,\
\end{equation}

\begin{equation}
V_{P}(12) = -\sum_{i=1}^{M}\sum_{j=1}^{M}\epsilon  
\exp \left[ -\frac{1}{2}{\left( \frac{r_{12}^{ij}}{\alpha} \right)}^{q} \right] \,\,\,\,\,\,\,\,\,\,\,\,\
\end{equation}

\noindent
Here $r_{12}$ is the distance between the center of mass of particles $1$ and $2$, $r_{12}^{ij}$ is the distance between patch $i$ on particle $1$ and patch $j$ on particle $2$, and $M=4$ is the number of patches per particle. The four patches are tetrahedrally distributed on the surface of the particles. Exponents in $V_{CM}(1,2)$ and $V_P(1,2)$ are taken as $p=200$ and $q=10$ to resemble the functional behaviours of a hard sphere and a square well interaction respectively. We selected $\alpha=0.12$ as the patch diameter to avoid having more than one bond per patch and $\epsilon=1.001$, for which the minimum of the attractive part of the potential energy in a bounded configuration is $u_0 \equiv \min {V_{P}(12)}=-1$. Temperature $T$ is measured in units of the potential well, taking Boltzmann's constant as $1$. The unit of time is $\sigma\sqrt{m/ |u_0|}$, being $m=1$ the mass of the particles. To integrate the equations of motion we used a Velocity Verlet algorithm with a fixed time step $\delta t=0.001$ using a modified Brownian thermostat which explicitly avoids unphysical decorrelations in the particle velocity ~\cite{John_valence,lorenzo-patchy}. The longest runs extended over $7 \cdot 10^9$ time steps for the lowest temperatures ($T =0.105$ and $T =0.1025$). All simulations were performed using the oxDNA simulation package running on GPUs~\cite{rovigatti_gpu_comparison}.\\

We investigated temperatures within the range $T \in [0.1025, 0.16]$, covering a slowing down of the dynamics of four orders of magnitude. The dynamics of this system exhibits Arrhenius behaviour at low temperature, a signature of strong glass-formers~\cite{soft_matter_tetrahedral}. The temperature at which the system enters into the Arrhenius regime is $T_A \cong 0.115$~\cite{rovigatti_molphys,soft_matter_tetrahedral} (Fig. 4 in the main text). We take as the \textit{onset} temperature for this system $T_0 = 0.16$ (Fig. 4 in the main text). Above this temperature the system is not completely percolated into an infinite cluster and the MSD does not present any detectable plateau~\cite{rovigatti_molphys}.\\

We evaluated dynamic observables by an average of $50$ independent simulations (500.000 individual particle trajectories) per each value of the temperature. The total amount of GPU time was approximately 2 months, which for this system corresponds to more than 3 years of CPU time on a high-end processor~\cite{rovigatti_gpu_comparison}.

\subsection{Estimation of $\lambda(t)$} 

We determined  $\lambda(t)$ via maximum likelihood assuming an exponential probability density of displacements from a given threshold $r_t$. The likelihood at time $t$ is given by:

\vspace{0.3cm}
\begin{equation}
\mathcal{L}_t = \prod_{i=1}^{M_t} \exp\left( -\frac{r_i-r_t}{\lambda_t} \right) / \lambda_t
\end{equation}

\vspace{0.3cm}
\noindent
where $M_t$ is the number of individual particle trajectories with a displacement $r_i$ larger than $r_t$ at time t.  We considered for convenience the minus logarithm of the likelihood,
\begin{equation}
l_t = -\log \mathcal{L}_t = \sum_{i=1}^{M_t} \frac{r_i-r_t}{\lambda_t} +M_t\log \lambda_t .
\end{equation}
\noindent
We then obtained the $\hat{\lambda}_t$ that minimizes $l_t$ with respect to $\lambda_t$:
\begin{equation}\label{eq:hatlambda}
\frac{\partial l_t}{\partial \lambda_t}\Bigr|_{\lambda_t = \hat{\lambda}_t} = 0 \,\ \Rightarrow \,\ \hat{\lambda}_t = \frac{1}{M_t}\sum_{i=1}^{M_t} (r_i-r_t)
\end{equation}

It is desirable to have the lowest possible threshold $r_t$ so that more data are included in the estimation while still having a statistically significant exponential tail. In order to achieve this, we used a standard iterative procedure for power-law tail evaluation~\cite{clauset2009power}. We started with a very large threshold $r_t$ and computed $\hat{\lambda}_t$ by means of Eq. 14. At this point, we performed a test of statistical significance: if the test is passed, \textit{i.e.} the data are well represented by the estimated exponential distribution, we decreased the threshold $r_t$ by $0.025\sigma$ (being $\sigma$ the particle diameter) and repeated the procedure; if the test fails, we stopped the procedure and took $\lambda(t) =\hat{\lambda}_t$.\\

The statistical test we used is the non-parametric Kolmogorov-Smirnov test~\cite{Kolmogorov-Smirnov}, which measures how extreme is the distance between the theoretical cumulative distribution under the null hypothesis (where the data are sampled from the theoretical distribution) and the empirical cumulative distribution. The distance between distributions is measured by the Kolmogorov-Smirnov statistics,

\begin{equation}
KS_{M_t}=\sup_{r_i(t)} |F_{null}(r_i(t))-F_{emp}(r_i(t))|
\end{equation}

\noindent
where $F_{null}(x)$ is the cumulative distribution under the null hypothesis (exponential distribution with exponential rate $\hat{\lambda}_t$) and $F_{emp}(x)$ the empirical cumulative distribution. The null hypothesis was accepted (rejected) when the $p$-value was greater (smaller) than $0.05$~\cite{Kolmogorov-Smirnov}.   

\subsection{Estimation of $\mu(t)$, $t_0$, and $\beta$} 

We define $y_t \equiv$ MSD($t$). Since each $y_t$ is obtained as an average over many independent, and equivalent, simulations (see previous sections), we assumed that it is normally distributed. We denote by $\bar{y}_{1,T}$ and $\sigma_{1,T}$ the mean and standard deviation of this distribution for a discrete time interval $\{t_1,...,t_T\}$. The corresponding likelihood is  

\begin{equation}
\mathcal{L}_{1,T} = \prod_{t=t_1}^{t_T} \exp\left( -\frac{(y_t-\bar{y}_{1,T})^2}{2\sigma_{1,T}^2} \right) / \sqrt{2\pi\sigma_{1,T}^2}
\end{equation}

As in the previous section, we considered the minus logarithm of the likelihood 

\begin{align}
\begin{split}
    l_{1,T} & = -\log \mathcal{L}_{1,T} = \\
    & = \sum_{t=t_1}^{t_T} \frac{(y_t-\bar{y}_{1,T})^2}{2\sigma_{1,T}^2} + \frac{T}{2}\log2\pi + T\log\sigma_{1,T}
\end{split}
\end{align}

We then assumed that $\bar{y}_{1,T}$ and $\sigma_{1,T}$ scale with the same power law exponent, $\mu_{1,T}$, within the discrete time interval $\{t_1,...,t_T\}$: $\bar{y}_{1,T}=a t^{\mu_{1,T}}$ and $\sigma_{1,T} = c t^{\mu_{1,T}}$, with $a$, $c$ $\in \mathbb R^+$. This assumption is valid when deviations from normality are small, where the variance $\sigma_{1,T}^2$ scales as $2y_t^2/(T-1)$ and, therefore, $\sigma_{1,T} \sim \bar{y}_{1,T}$. This results in

\begin{align}
\begin{split}
l_{1,T} & = \sum_{t=t_1}^{t_T} \frac{(y_t-a t^{\mu_{1,T}})^2}{2c^2t^{2\mu_{1,T}}} + \\
& + \frac{T}{2}\log2\pi + T\log c + \mu_{1,T}T\log t
\end{split}
\end{align}

\vspace{0.3cm}
This function was minimized with respect $\mu_{1,T}$, $a$, and $c$ via the L-BFGS-B algorithm implemented in scipy.optimize.minimize~\cite{scipy}, and following the procedure detailed in~\cite{Leitao-scaling}.\\ 

In Eq. 18 $\mu_{1,T}$ corresponds to the discrete window given by $\{t_1,...,t_T\}$, which was chosen to cover one decade in time. To associate $\mu_{1,T}$ to a specific time, we chose the middle time in the sequence, \textit{i.e.} $t_{(1+T)/2}$, being $T$ an odd natural number. In this way, the value $t_0$ associated to the entrance to the diffusive regime is the middle time in that sequence $\{t_1,...,t_T\}$ for which $\mu_{1,T} = \mu(t_{(1+T)/2}) \equiv \mu(t_0) = 0.95$. The reason for this choice is that $\mu=1$ is an asymptotic limit ($t\rightarrow \infty$) which is approached from below but never reached. The criterion $\mu = 0.95$ ensures that the exponent $\mu$ is compatible with  $\mu=1$ up to our measured standard error, which is equal to  0.05 on average.\\

Finally, to estimate $\beta$ we considered time windows of one decade starting at $t_0$ (see Figs. 2 and 3), with the exception of those simulations where the Brownian yet non-Gaussian regime does not cover one decade in time (see for instance T = 0.1025 in Fig. 3). Then we fitted $\lambda(t \geq t_0)$ by a least-squares regression assuming a power law $\lambda(t \geq t_0) = a_1 t^{\beta} + a_2$, with $a_1, a_2 \in \mathbb R^+$ as free parameters.   

\section*{Acknowledgements}

We thank Walter Kob for fruitful exchanges about this work and Lorenzo Rovigatti for his assistance with the tetrahedral system simulations. We also thank the anonymous reviewers for their careful reading of our manuscript and their insightful comments and suggestions. A.V.C. acknowledges partial financial support from the Deutsche Forschungsgemeinschaft (DFG Grant ME 1535/7-1). S.R.-V. acknowledges support from the European Commission through the Marie Skłodowska-Curie Individual Fellowship 840195-ARIADNE. 

\bibliographystyle{apsrev}

\line(2,0){225}
\clearpage
\section*{Appendix A: scaling in a 4D Lennard-Jones system}

In this Appendix we provide evidence of the scaling behavior $\lambda (t) \sim t^{1/d}$ for the Lennard-Jones system in $d=4$ from the analysis of the exponential tails of $P(r;t) = \phi_4 G(r;t)$, where $\phi_4 = 2\pi^2$ (see Eq. (2)). We performed simulations for a 4D-Kob-Andersen binary mixture with the same interaction parameters and units defined in the Methods for $d=2,3$. We simulated a system with $N= 10^4$ particles (five times the number of particles used in Ref.~\cite{kob-andersen_2D}), with $N_A=6500$ and $N_B=3500$. The linear size of the simulation box was $L=8.493581$. The total simulation time was $1.6 \cdot  10^5$ and the total number of independent trajectories of A particles analyzed $1.3\cdot 10^4$. The composition and number density is the same as in Ref.~\cite{kob-andersen_2D} for $d=4$. We investigated two temperatures, using an integration time step $\delta t^{4D} = 0.02$. The higher $T$ is an intermediate temperature which presents an incipient plateau at intermediate times for the MSD, whereas the lower temperature presents a marked plateau at intermediate times (Fig.~6a)). With the chosen temperatures we enclose a wide range from the beginning of the glassy dynamics to the inner supercooled regime, leaving a more detailed study of the functional behavior of $\beta(T)$ in 4D to future studies.\\

As in Fig. 2, we fit in Fig.~6b the scaling of $\lambda(t)$ for the 4D-Lennard-Jones system within a time window where $t \geq t_0$ for the two temperatures investigated. This scaling is compatible with $\lambda(t) \sim t^{\beta}$. The fitted values of $\beta$ were $\beta (T=0.750)=0.29 \pm 0.05$ and $\beta (T=0.675)=0.28\pm 0.04$. Both of these values are compatible with the theoretical value $\beta= 1/4$ as predicted by the argument embodied in Eqs.~(4) to (6). \\

Due computer limitations, our statistics for the 4D-Lennard-Jones system resulted in a time window smaller than one decade. For $t \gtrsim 4 t_0$ the data become noisy and the exponential tails were in general non accepted by the Kolmogorov-Smirnov test within a sufficiently large $r$-range. To verify the reliability of the results, we performed an additional chi-squared goodness-of-fit test using three families of fundamental increasingly monotonic functions, all of them with two free parameters. First we fitted $\lambda(t)$ within the explored time window using the power law predicted by our scaling argument $\alpha t^\beta$, with $\alpha > 0$ and $\beta > 0$ as free parameters. Second we fitted $\lambda(t)$ using the family $\alpha e^{\beta t}$, exactly for the same time window and the same number of points (50 in the plot). Finally, we fitted $\lambda(t)$ using the family $\alpha +\beta \log(t)$, again for the same time window and number of points. Thus, for the same number of degrees of freedom, the goodness of the test for the power law was greater than $99\%$ for the two temperatures investigated, whereas the goodness of the exponential family was less than $30\%$ and the logarithmic family less than $1\%$. This statistical test supports the power law behavior and the value $\beta \cong 1/4$ predicted by our scaling argument.

\begin{figure}[H]
\center
\includegraphics[width=0.46\textwidth]{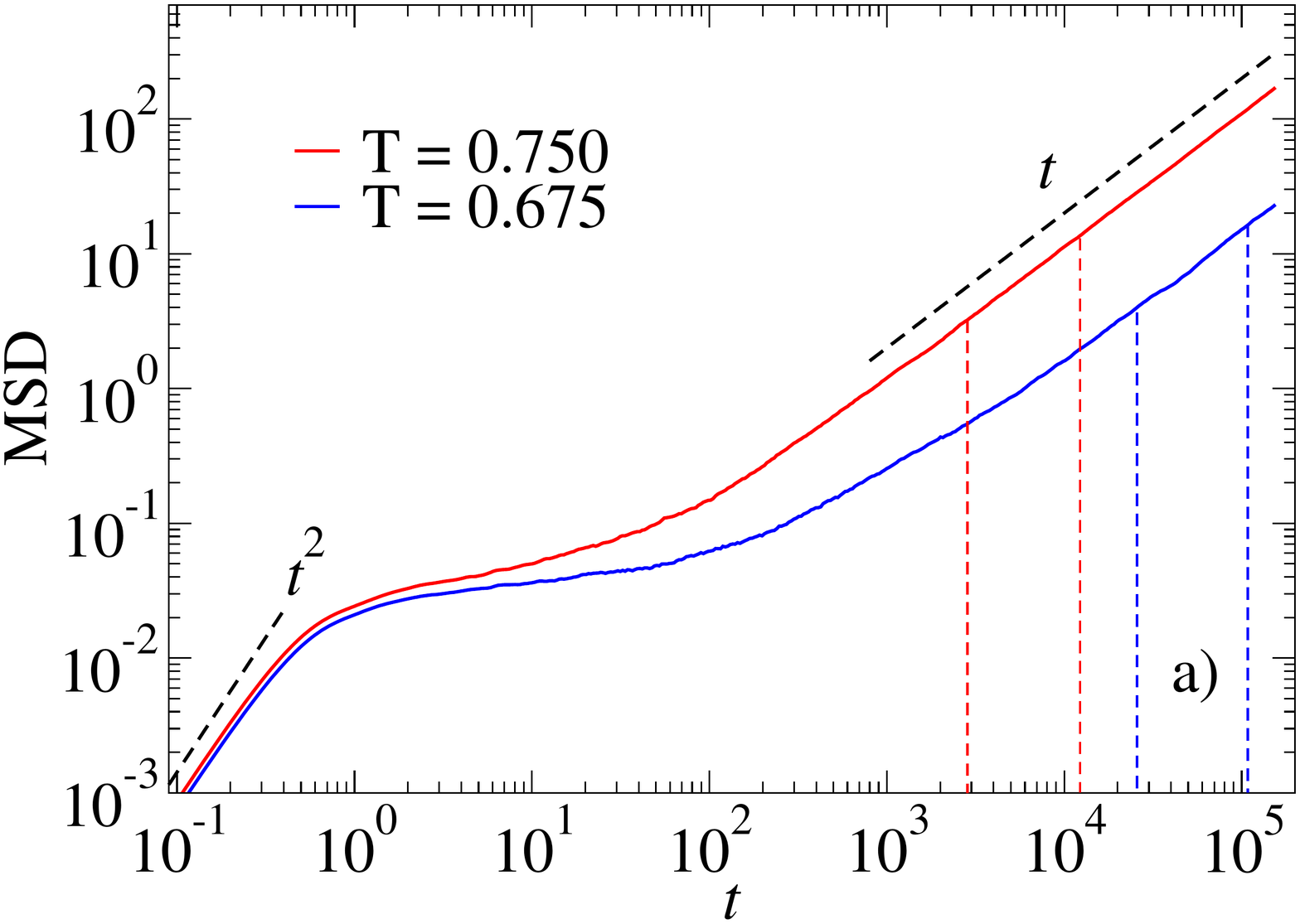}
\includegraphics[width=0.49\textwidth]{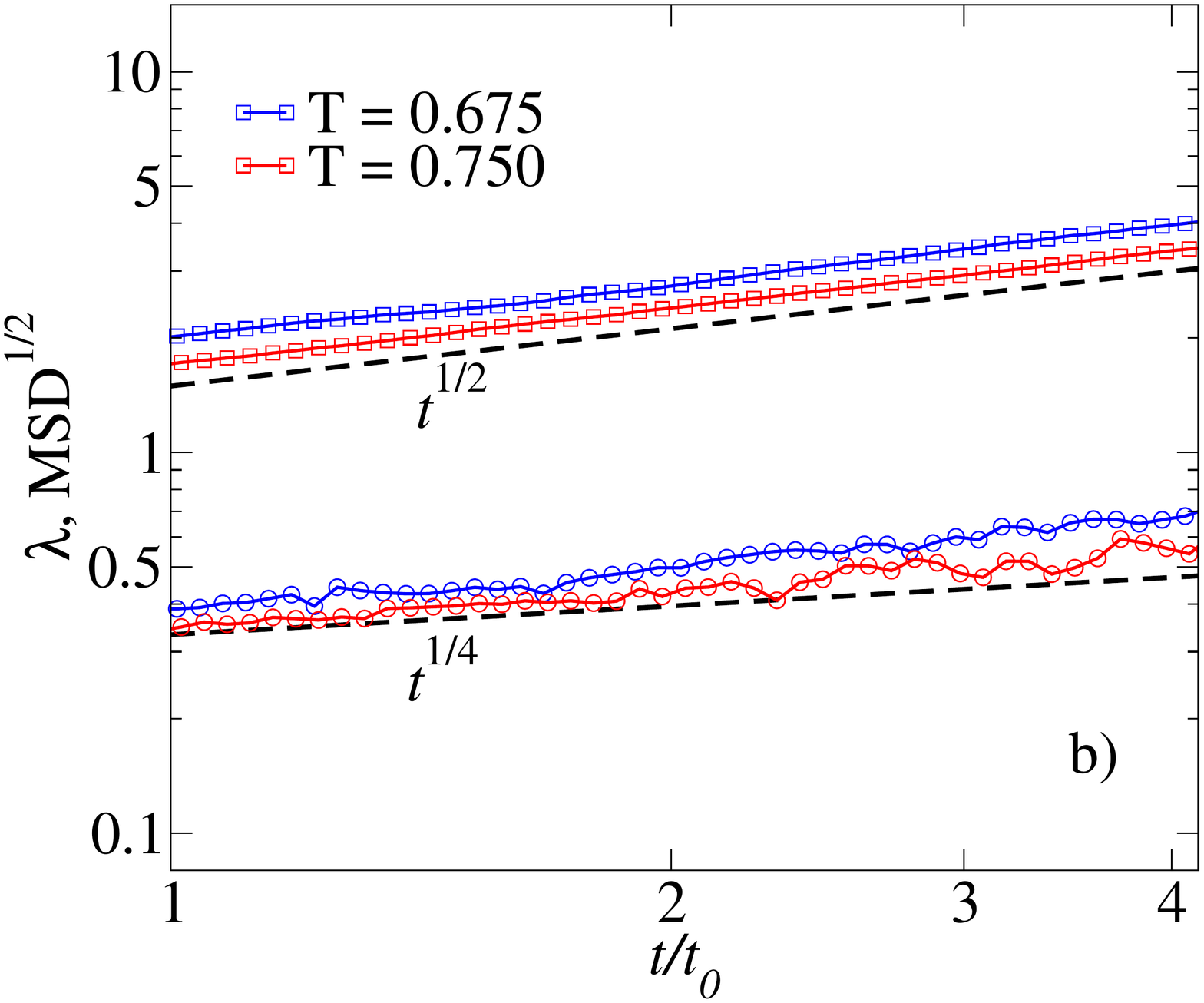}
\caption{\textbf{MSD and Brownian yet non-Gaussian regime for the 4D-Kob-Andersen binary mixture.} a) Double log plot of the MSD for the two temperatures investigated, covering the whole time regime from ballistic to diffusive motion. We mark by vertical dashed lines the time windows corresponding to the analyzed Brownian yet non-Gaussian regime for each temperature; b) Double log plot of MSD$^{1/2}$ (squares) and $\lambda$ (circles) as a function of time for the 4D-Kob-Andersen binary mixture at different temperatures: $T =$ 0.750 and 0.675. Here  $t_0$ is the temperature-dependent time at which the system reaches a $\mu$ value compatible with 1 within numerical uncertainty, \textit{i.e.} MSD($t \geq t_0$) $\sim t^{\mu \, \cong \,  1}$ (see Methods). The analyzed time windows correspond to the time regimes signaled by vertical dashed lines in a). Dashed lines representing power law behaviors in a) and b) serve as a reference.}
\center
\label{4D} 
\end{figure}

\section*{Appendix B: finite size effects in a 2D Lennard-Jones system}

In this Appendix we explore whether finite size effects influence our results for the 2D-Lennard-Jones system. We compare the system presented in Methods with an equivalent larger system (same composition, number density, and interaction parameters). We choose an intermediate temperature $T=0.6$ already presented in Fig. 2b. The larger system has a total number of particles $N=4 \cdot 10^4$, with $N_A=2.6 \cdot 10^4$ and $N_B=1.4 \cdot 10^4$, {\em i.e.} four times the number of particles in the system described in Methods. The linear size of the simulation box was $L=185.56246$ (twice the size in the system described in  Methods). The total simulation time was $2 \cdot  10^4$ and the total number of independent trajectories of A particles $6\cdot 10^5$ (the same number of trajectories in the  system described in  Methods at $T=0.6$).\\

\begin{figure}[H]
\center
\includegraphics[width=0.46\textwidth]{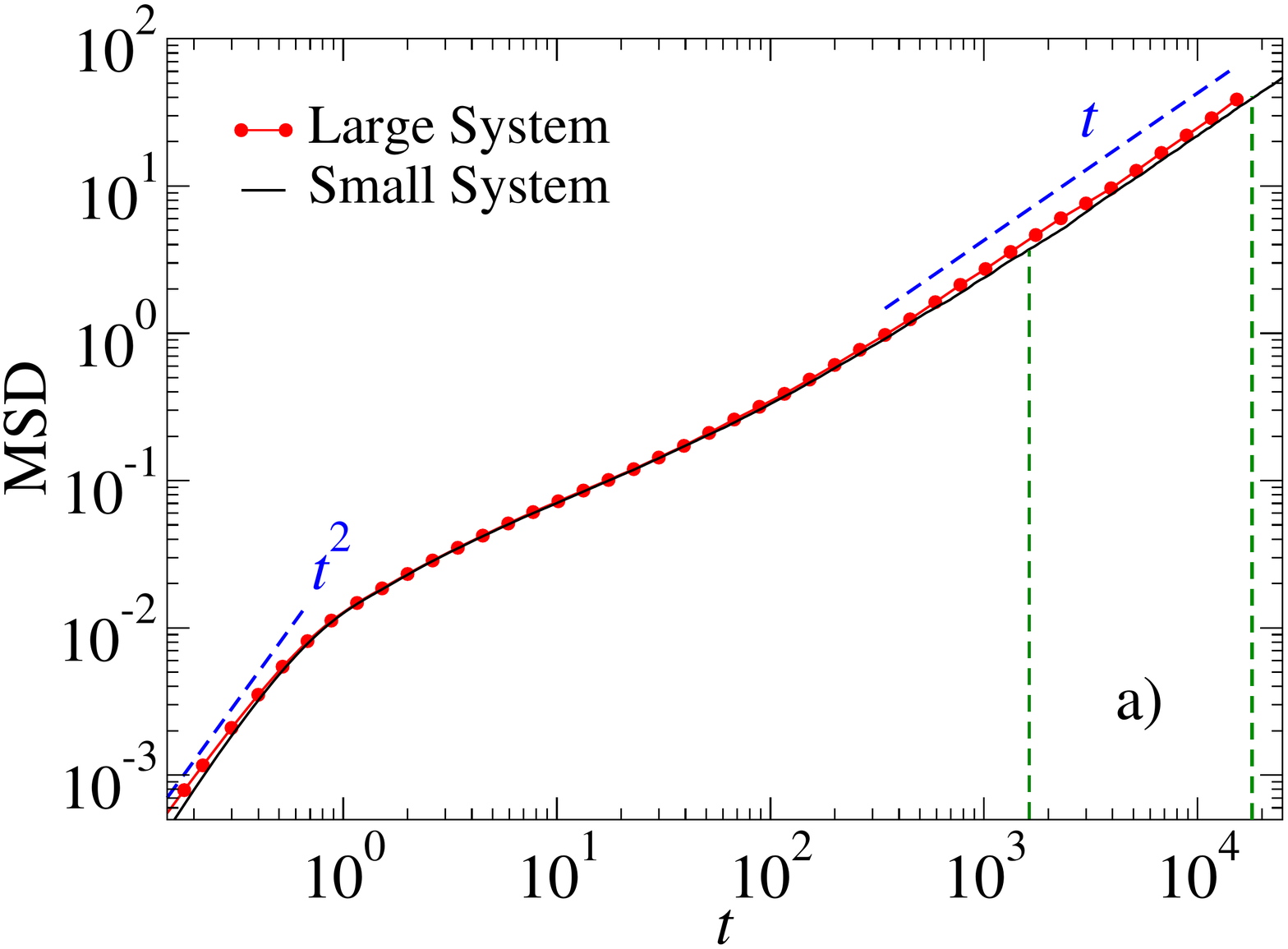}
\includegraphics[width=0.46\textwidth]{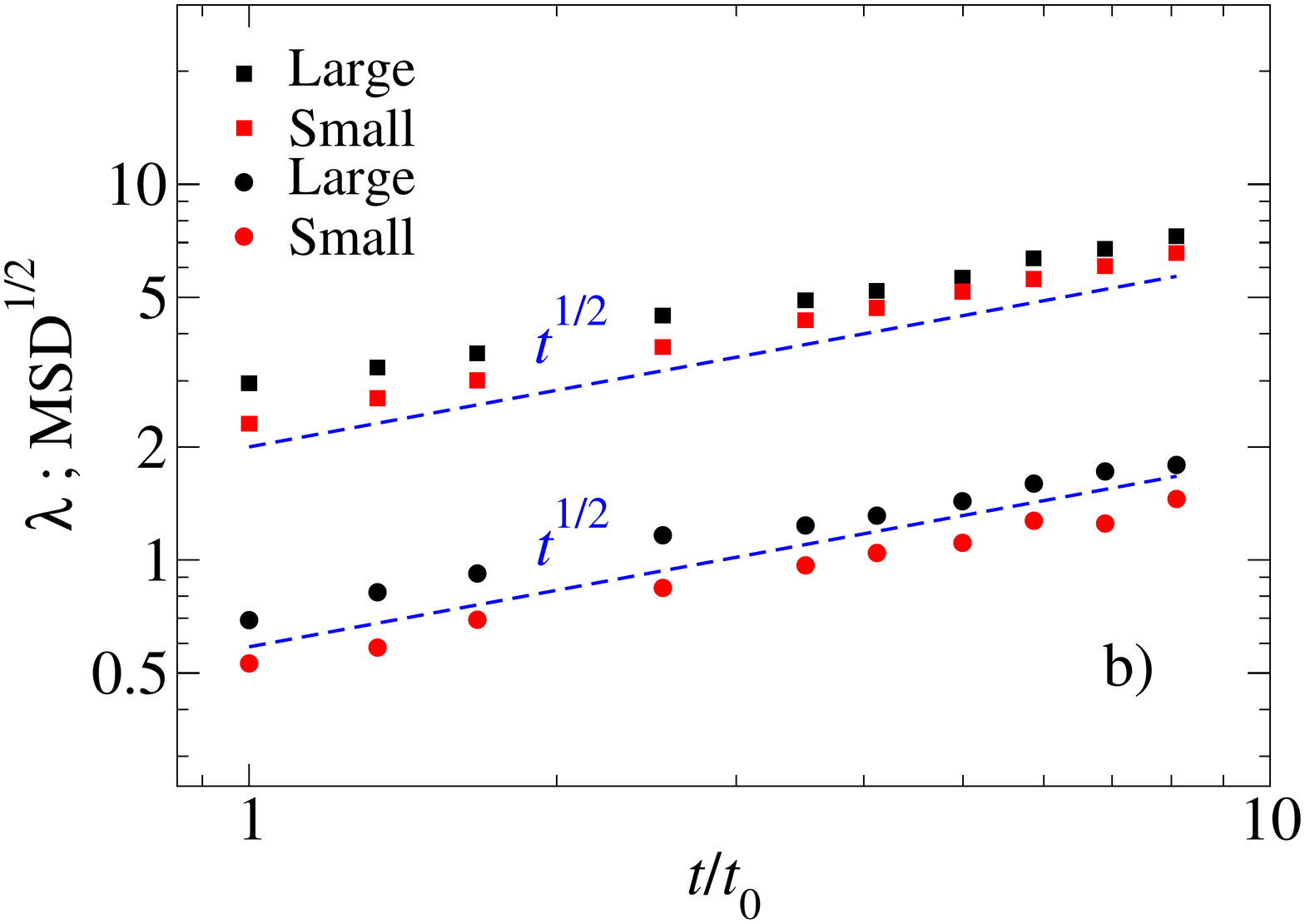}
\caption{\textbf{MSD and Brownian yet non-Gaussian regime for two 2D-Kob-Andersen binary mixtures of different size.} a) Double log plot of the MSD at $T=0.6$ for the large and small system, covering the whole time regime from ballistic to diffusive motion. We mark by vertical dashed lines the time window corresponding to the analyzed Brownian yet non-Gaussian regime; b) Double log plot of MSD$^{1/2}$ (squares) and $\lambda$ (circles) as a function of time for the large and small system at $T = 0.6$. Here  $t_0$ is the temperature-dependent time at which the systems reached a $\mu$ value compatible with 1 within numerical uncertainty, \textit{i.e.} MSD($t \geq t_0$) $\sim t^{\mu \, \cong \,  1}$ (see Methods). The analyzed time window corresponds to the time regime signaled by vertical dashed lines in a). Dashed lines representing power law behaviors in a) and b) serve as a reference.}
\center
\label{4D} 
\end{figure}

In Figure 7a we show a comparison of the MSD for the two systems. We only appreciate a tiny difference in the diffusive regime, where the larger system seems to present a slightly greater diffusion coefficient than the small system. Thus, as shown in Fig. 7b, while the two systems arrive at the diffusive regime almost at the same time, there is a small shift in both the MSD and $\lambda$ within the Brownian yet non-Gaussian regime. In any case, when comparing the dynamics of the two systems within this regime we see that both systems present a common power law behavior compatible with $\beta=1/2$, as already documented for the small system in Section~\ref{sec:exponentialtail}. The fitted values were $\beta (large) = 0.45 \pm 0.05$ and $\beta (small) = 0.47 \pm 0.04$. This comparison confirms that finite size effects negligibly affect our main result for 2D-Lennard-Jones systems: the power law dependence $\lambda(t) \sim t^{1/2}$ in the Brownian yet non-Gaussian regime.

\begin{figure}[H]
\center
\includegraphics[width=0.48\textwidth]{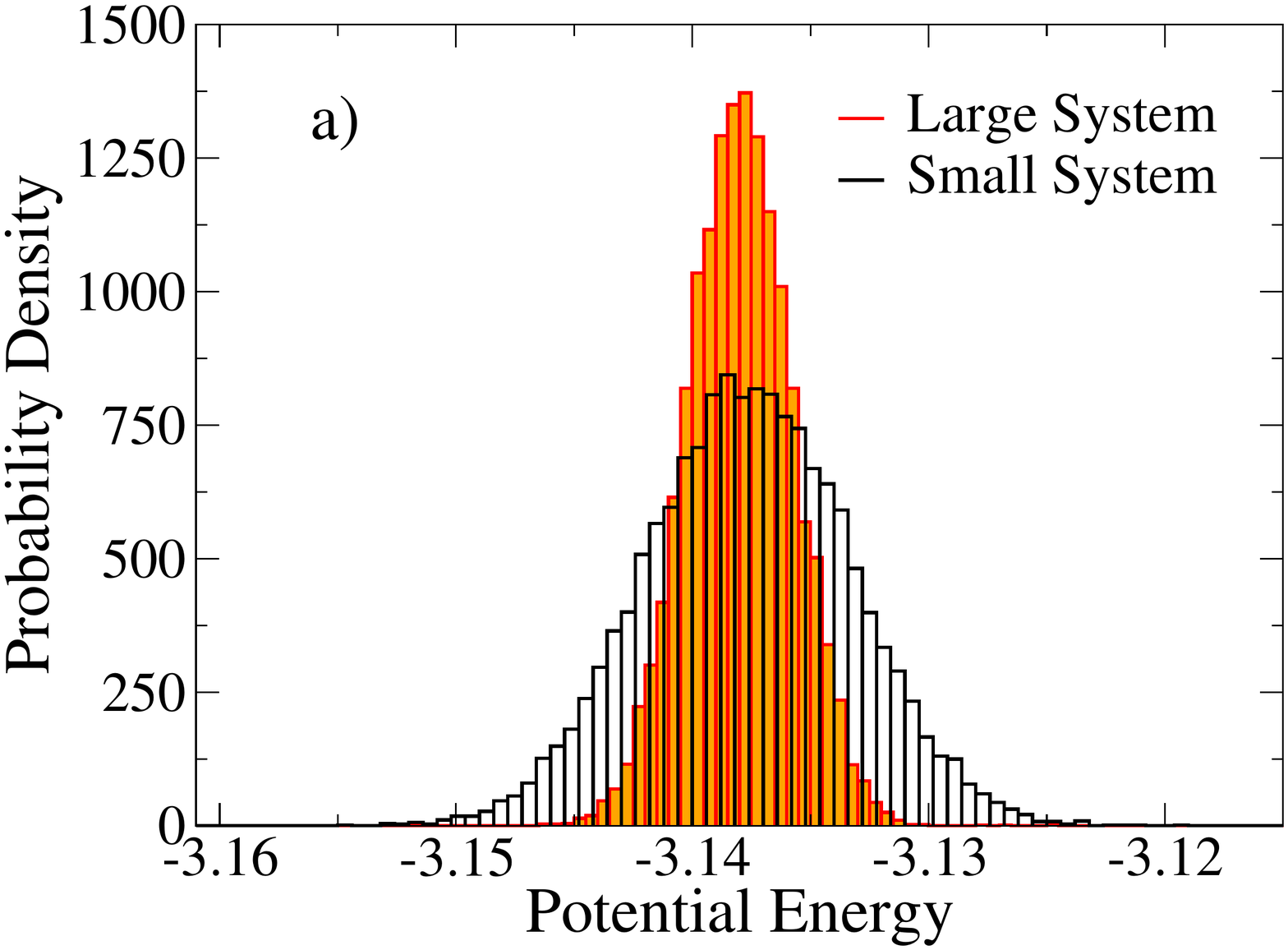}
\includegraphics[width=0.48\textwidth]{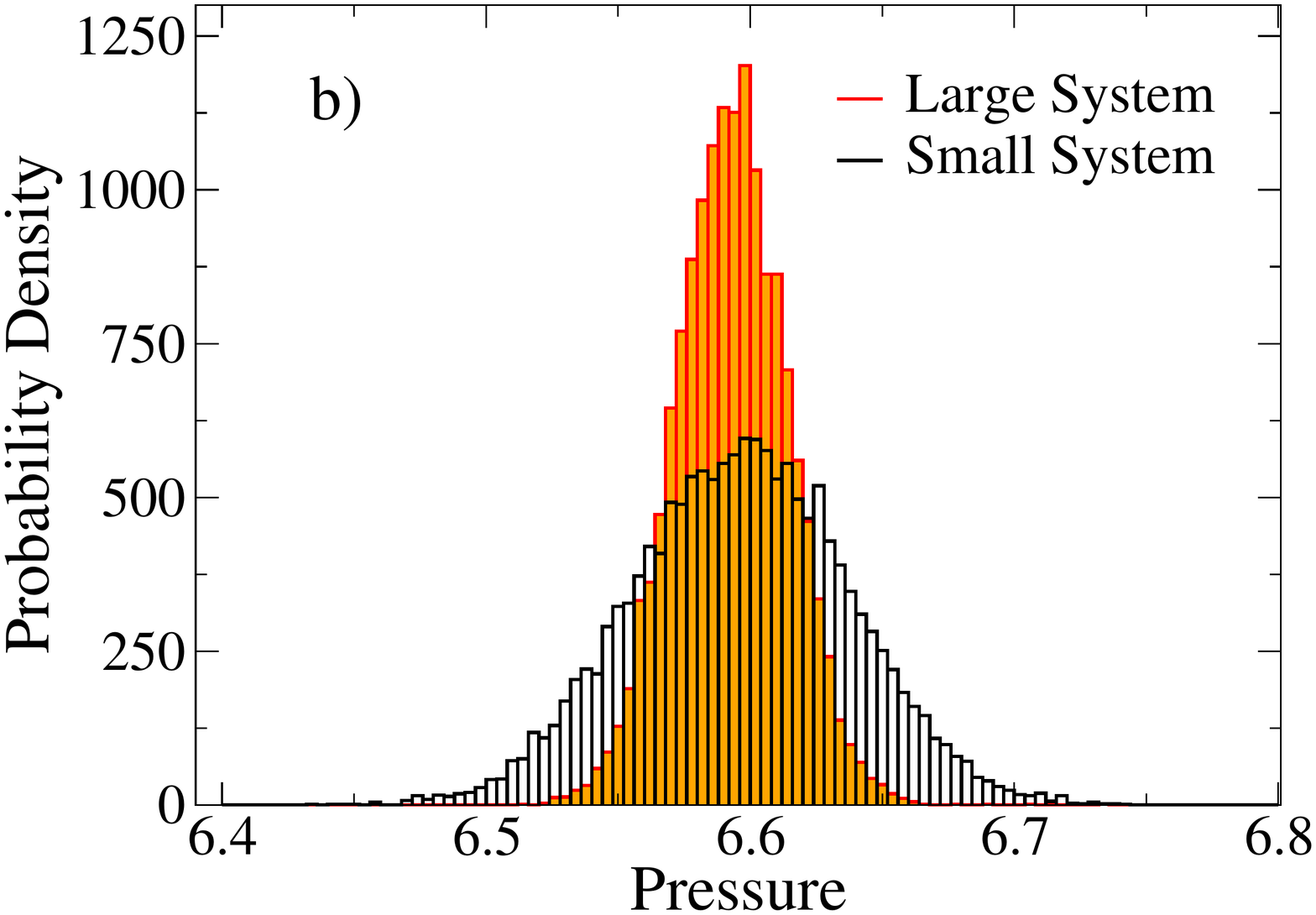}
\caption{\textbf{Potential energy and pressure for two 2D-Kob-Andersen binary mixtures of different size.} a) Probability density of the Potential energy per particle for the large and small system at $T = 0.6$ monitored every 20 time units for a total time $2 \cdot 10^4$; b) Probability density of Pressure for the large and small system at $T = 0.6$ monitored as in a).}
\center
\label{4D} 
\end{figure}

We perform an additional comparison for two static observables: potential energy and pressure. The probability densities of the potential energy for the small and large systems are peaked at about the same mean value: $-3.139$ for the large system and $-3.140$ for the small system, see Fig.~8a. The difference between the two distributions is in their standard deviations: $\delta_E(large) = 0.007$ and $\delta_E(small) = 0.015$. As expected by the law of large numbers, the standard deviation decreases at increasing the number of particles as $\delta_E(small)/\delta_E(large) \approx\sqrt{N_{large}/N_{small}} = 2 $. In Fig. 8b we show similar results for the pressure: a common mean value of $6.60$ for the two systems and a factor two in the standard deviations, $\delta_P(large) = 0.06$ and $\delta_P(small) = 0.12$.  

\section*{Appendix C: testing the scaling argument}

We test the scaling argument embodied in Eqs. (4) to (6) of the  Main Text. We first consider the 3D Lennard-Jones and 3D tetrahedral gel systems at their respective lowest temperatures. We wish in particular to show why the scaling argument fails for the tetrahedral system at low $T$, given as a result a $\beta$ exponent significantly smaller than $1/3$ (see Figure 4 in the Main Text).\\

We first focus on $P(r;t)$ at different times within the diffusive yet non-Gaussian regime ($t \geq t_0$), see Figure 9. The different exponential tails extend up to $r=0$ (straight lines in Fig. 9) and all intersect approximately at a common value at $r=0$. Therefore, Eq. (5), which assumes a non-evolving $\rho_0$, is well satisfied for both systems. At short distances, $P(r;t)$ decreases with time for the 3D Lennard-Jones. In contrast, $P(r;t)$ for the tetrahedral gelling system presents two favorable distances (inset in Fig. 9, right) at which the probability increases with time, at least within the time regime investigated. One distance corresponds to the average first neighbour distance ($r\cong 1$) and the other one corresponds to the tetrahedral neighbour distance ($r\cong 1.7$). This result is consistent with the underlying amorphous tetrahedral structure at low $T$ which still manifests itself, despite many particles have already jumped. In other words, some particles have moved to a position at time $t$ which was occupied by another particle at $t=0$.\\

Our next step is to test Eq. (4) of the Main Text, which establishes a linear relation between the number of hoppers within the diffusive yet non-Gaussian regime and time: $N_h(t) \approx \omega N t \sim t$. We estimate $N_h(t)$ by integrating for both systems the exponential tails shown in Fig. 9. This is presented in Figures 10a and b. While the 3D Lennard-Jones system satisfies Eq. (4) for the lowest $T$ investigated ($N_h(t) \sim t$), the tetrahedral gelling system presents a power law behavior compatible with $N_h(t) \sim t^{1/2}$, in violation of Eq. (4).\\ 

Two different, but non-mutually exclusive, mechanisms  could explain why Eq. (4) fails for the tetrahedral system. First, the tetrahedral system maintains a underlying structure at times greater than $t_0$. This structure is revealed by the two peaks in the inset of Fig. 9 (right). This means that some particles that have already jumped to a distance corresponding to one of these peaks, could in principle jump back to their original positions, thus abandoning the exponential tail. Such a mechanism would slow down the growth of  $N_h(t)$ with time. Second, we also speculate that the tetrahedral system could produce hoppers intermittently in time. In this way, the tetrahedral system would present avalanches in the  production of hoppers alternated with  periods of scarce production. As a result, the empirical total production of hoppers given by $N_h(t)$ would not be linear before reaching the final Gaussian regime, at which  all the particles have already jumped. We leave a detailed test of these hypotheses for future studies. \\

Finally, we directly test Eq. (6) of the Main Text by plotting $N_h$ versus $\lambda$ in Figures 10c and d. We obtain $N_h(\lambda)$ from $N_h(t)$ at a fixed time $t$ and take the $\lambda$ value corresponding to this time $t$. In this case, the behavior of both systems is compatible with Eq. (6): $N_h(\lambda) \sim \lambda^{d=3}$. In summary, the results of Fig. 10 show: $N_h(t) \sim t$ and  $N_h(\lambda) \sim \lambda^3$ for the 3D Lennard-Jones system; and $N_h(t) \sim t^{1/2}$ and  $N_h(\lambda) \sim \lambda^3$ for the tetrahedral system. This results in $\lambda \sim t^{\beta=1/3}$ for the 3D Lennard-Jones system; and $\lambda \sim t^{\beta=1/6}$ for the tetrahedral system (see the $\beta$ exponent at lowest temperature for the 3D Lennard-Jones and the tetrahedral systems in Fig. 4 in the Main Text).\\

For completeness, we test our scaling argument in the 2D Lennard-Jones system at $T=0.36$. The common intersection of the exponential tails is shown in Fig. 11a, the predicted behavior $N_h(t) \sim t$ is presented in Fig. 11b,  and the behavior $N_h(\lambda) \sim \lambda^2$ in Fig. 11c. All these partial results lead to $\lambda \sim t^{1/2}$, as predicted by the scaling argument (Eqs. (4) to (6)).      

\newpage

\begin{figure*}
\centering
\captionsetup{justification=centering,margin=0.5cm}
\includegraphics[width=0.43\linewidth]{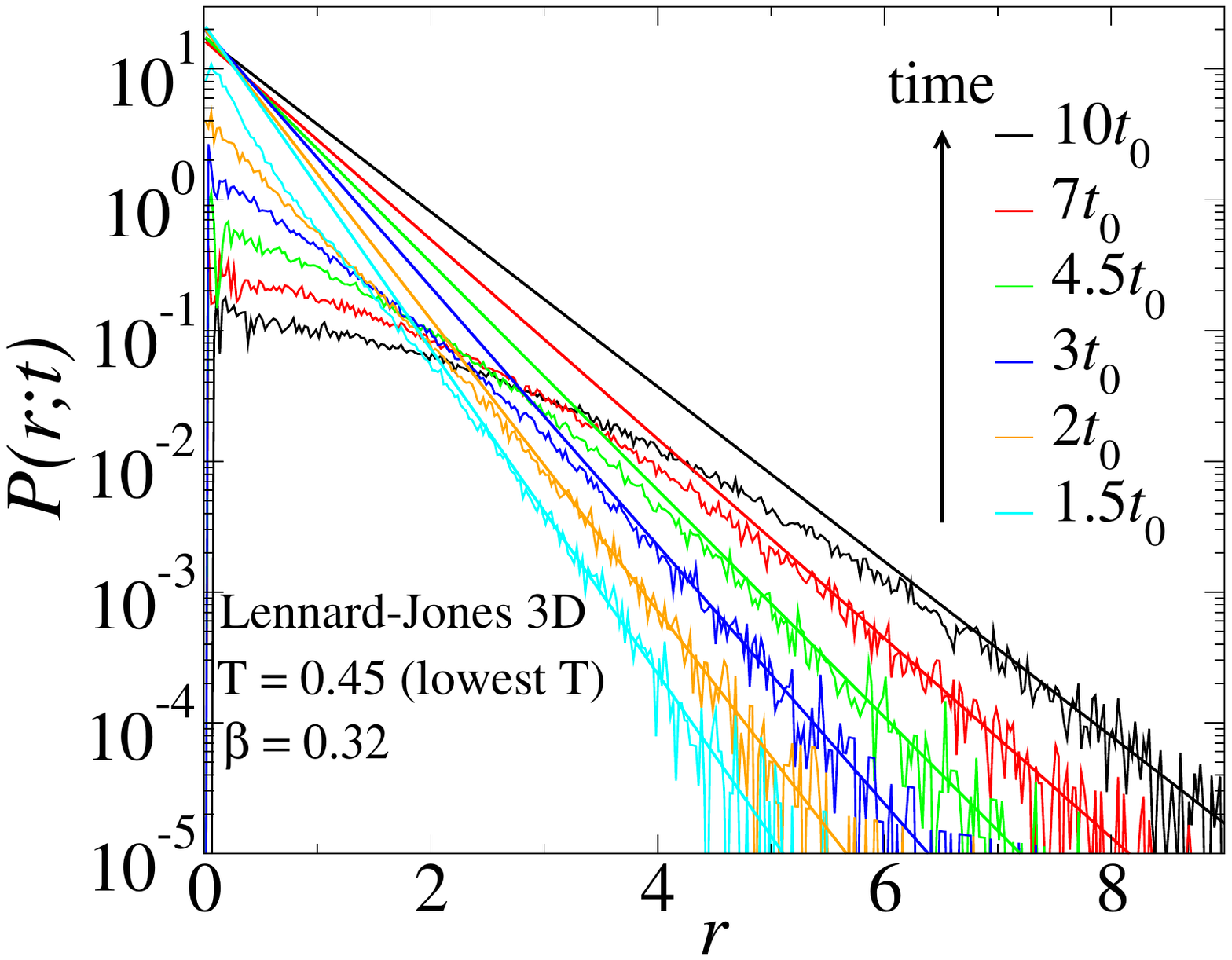}
\includegraphics[width=0.45\linewidth]{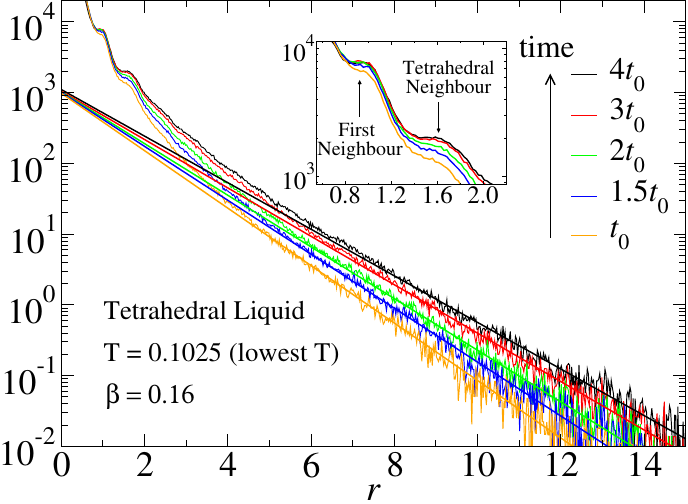}
\caption{\underline{Left}: $P(r;t)$ at different times $t$ ($> t_0$) for the 3D-Kob-Andersen binary mixture at the lowest temperature investigated ($T = 0.45$). \underline{Right}: $P(r;t)$ at different times $t$ ($\geq t_0$) for the tetrahedral gelling system at the lowest temperature investigated ($T = 0.1025$). The inset in the right figure shows a detail of $P(r;t)$ at distances corresponding to the first and tetrahedral neighbours. Straight lines for both systems correspond to the exponential tails of the different $P(r;t)$, extended to $r=0$.}
\end{figure*}

\begin{figure*}
\centering
\captionsetup{justification=centering,margin=0.5cm}
\includegraphics[width=0.44\linewidth]{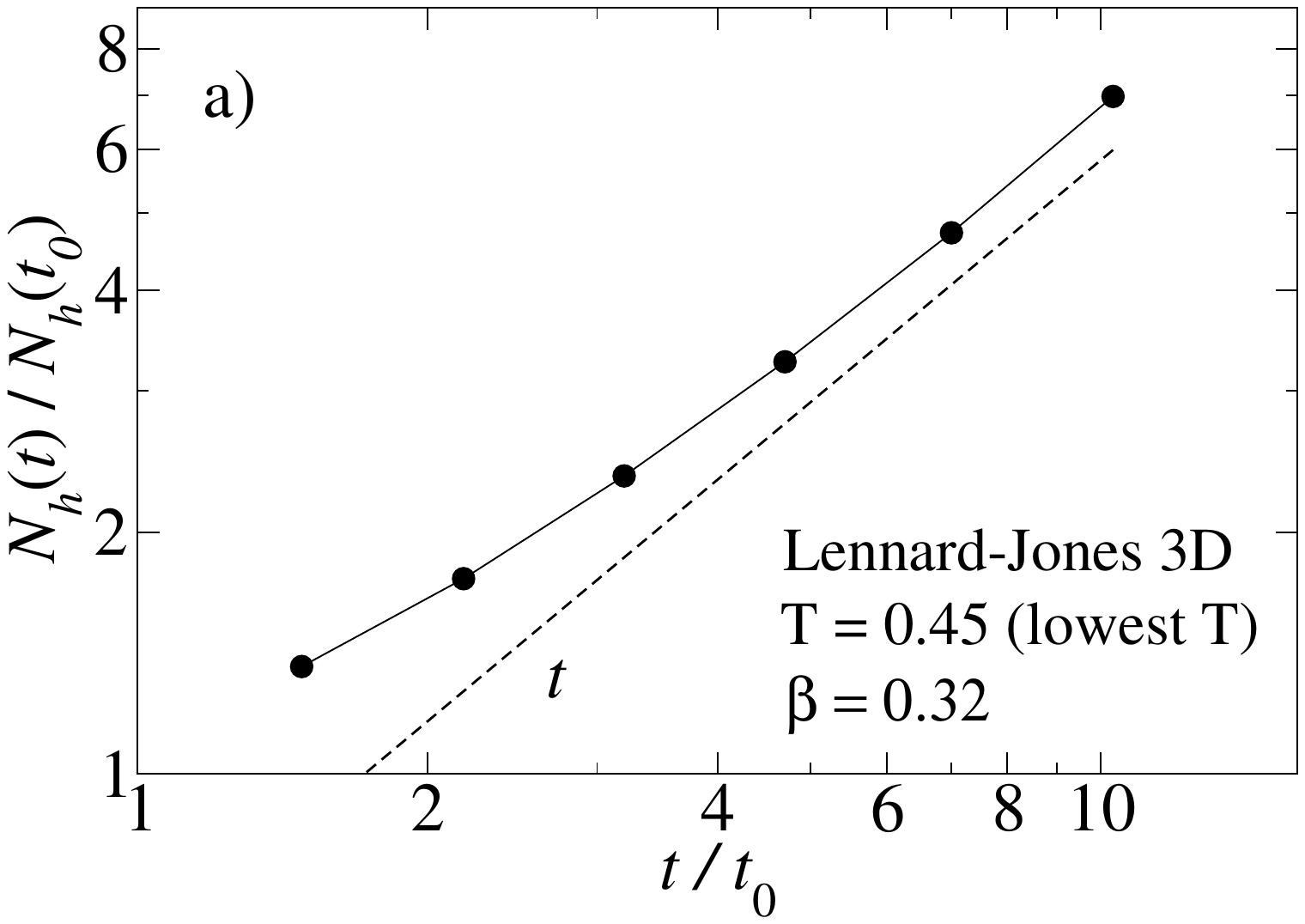}
\includegraphics[width=0.44\linewidth]{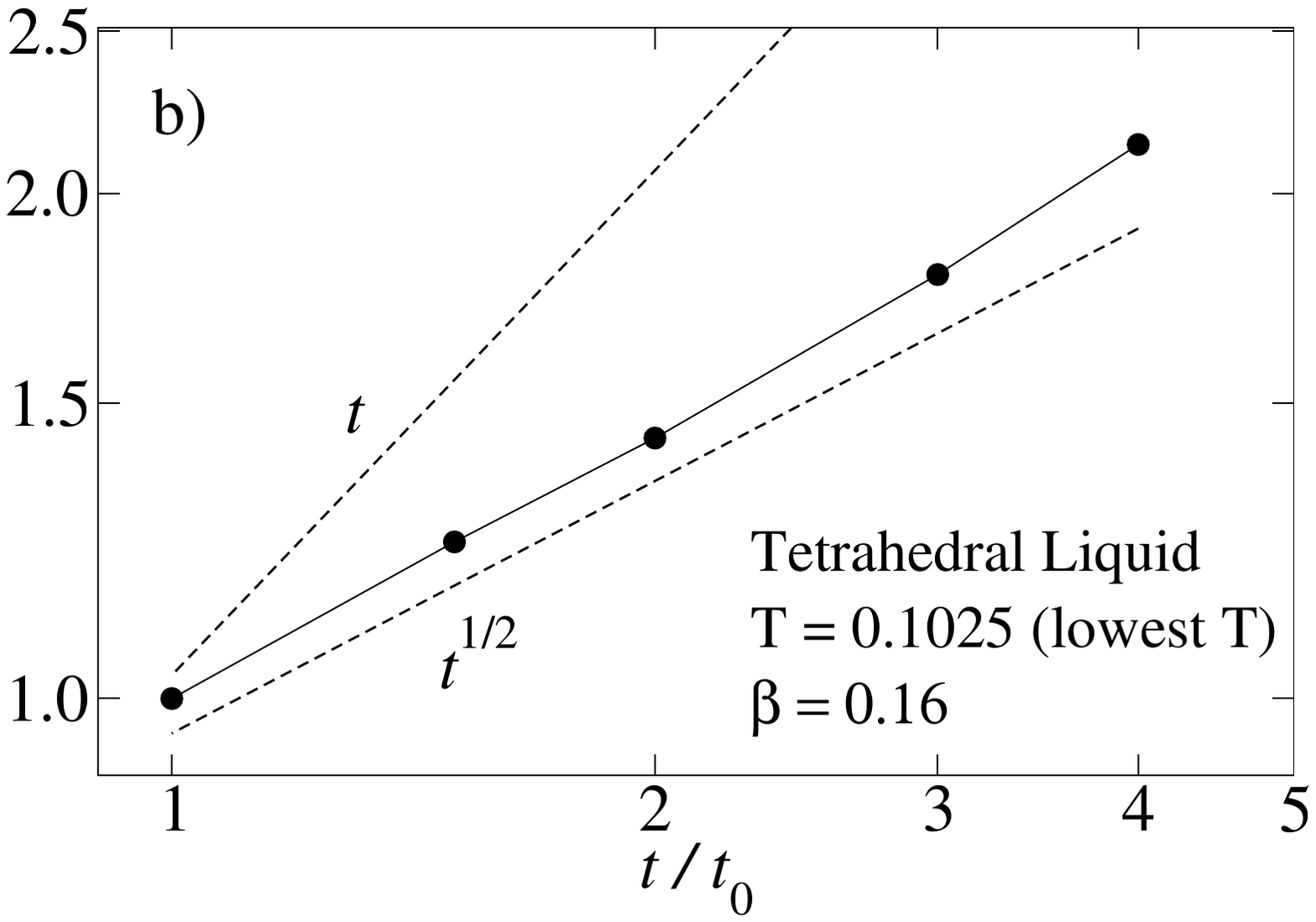}\\
\includegraphics[width=0.44\linewidth]{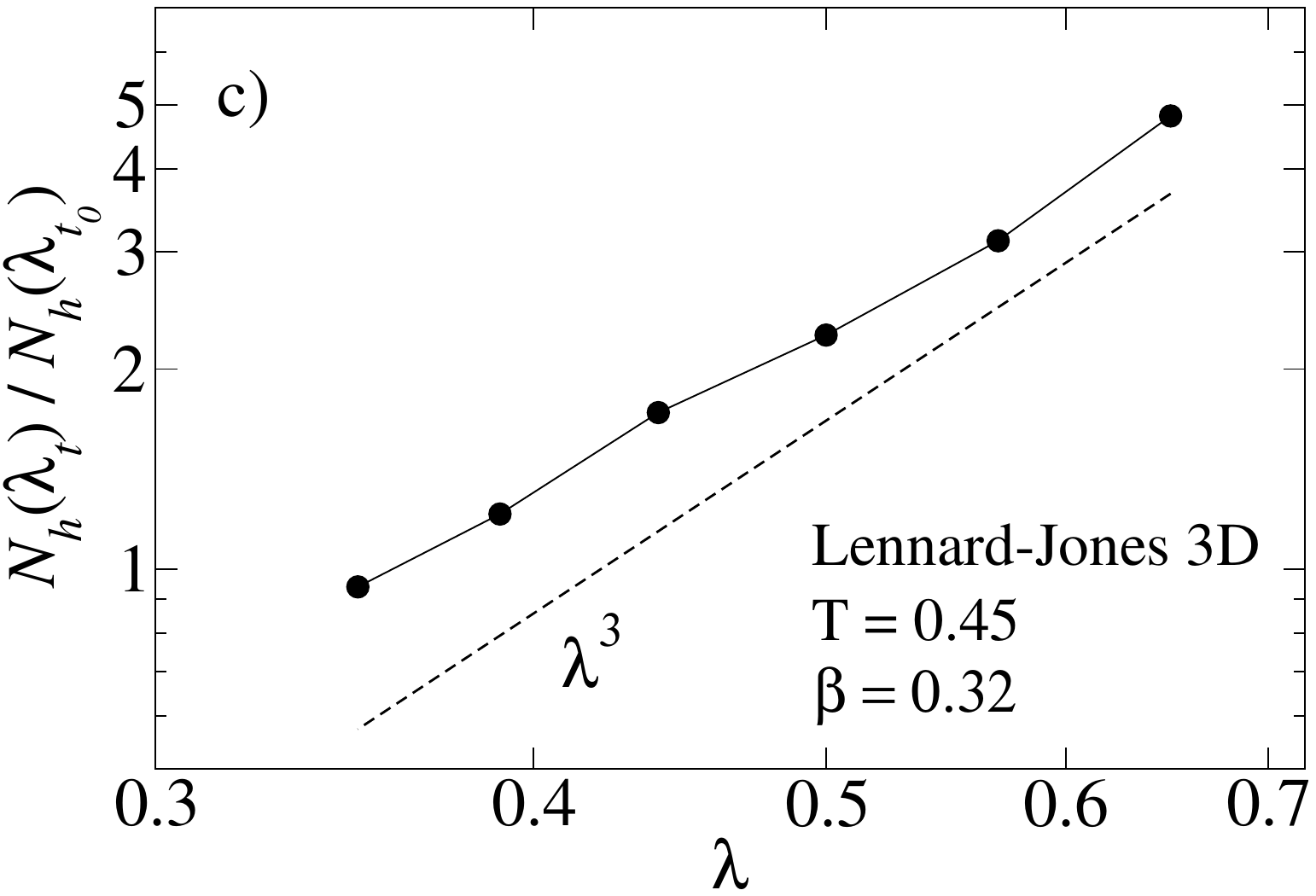}
\includegraphics[width=0.44\linewidth]{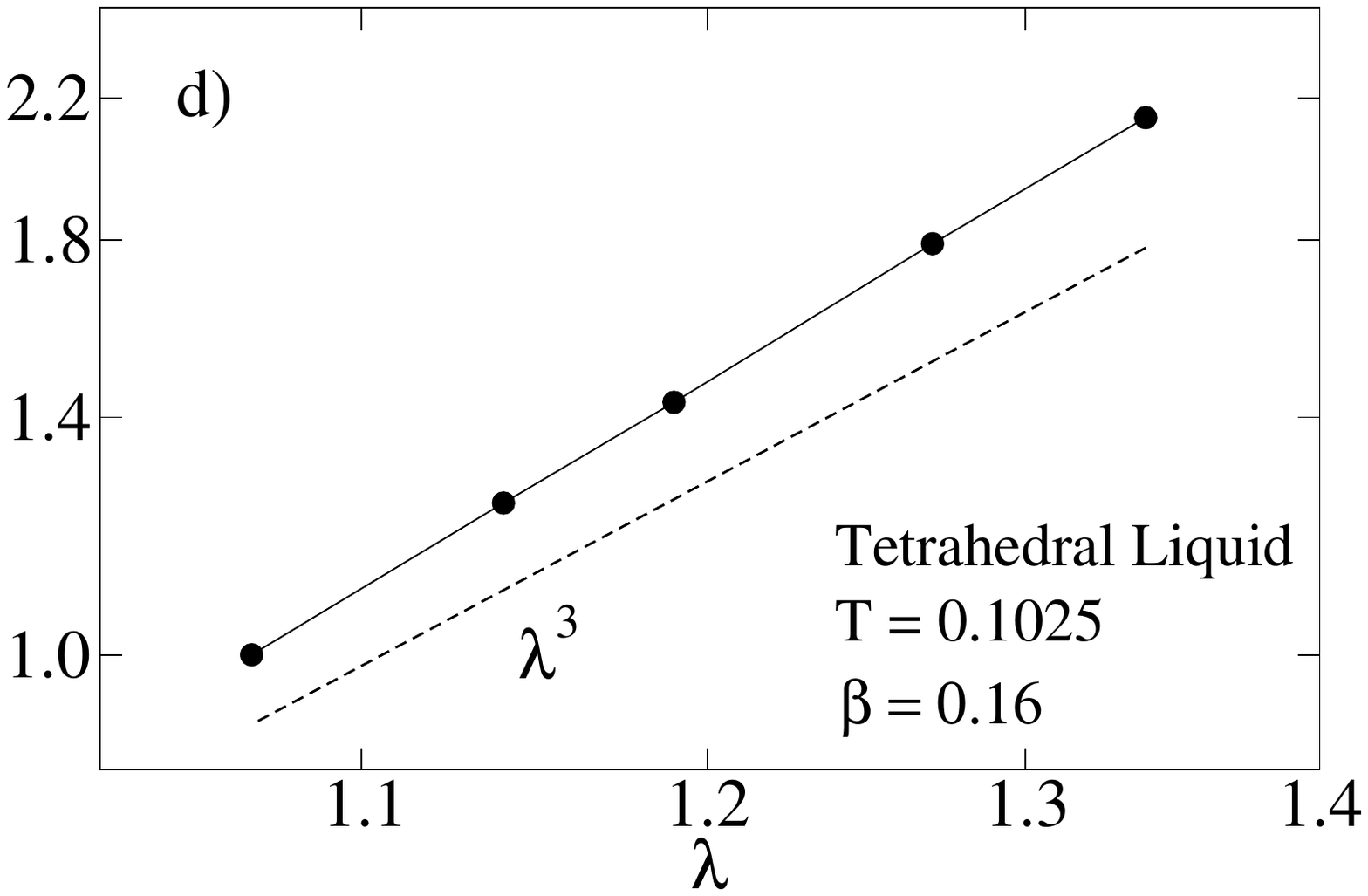}
\caption{$N_h(t)$ normalized by $N_h(t_0)$ for the 3D-Kob-Andersen binary mixture at $T = 0.45$ (a)) and the tetrahedral gelling system at $T=0.1025$ (b)), both obtained by integrating the exponential tails shown in Fig. 9. $N_h(\lambda)$ normalized by $N_h(\lambda_{t_0})$ for the 3D-Kob-Andersen binary mixture at $T = 0.45$ (c)) and the tetrahedral gelling system at $T=0.1025$ (d)). Dashed lines in all the figures are different power laws, shown as references.}
\end{figure*}

\begin{figure*}
\centering
\captionsetup{justification=centering,margin=0.5cm}
\includegraphics[width=0.6\linewidth]{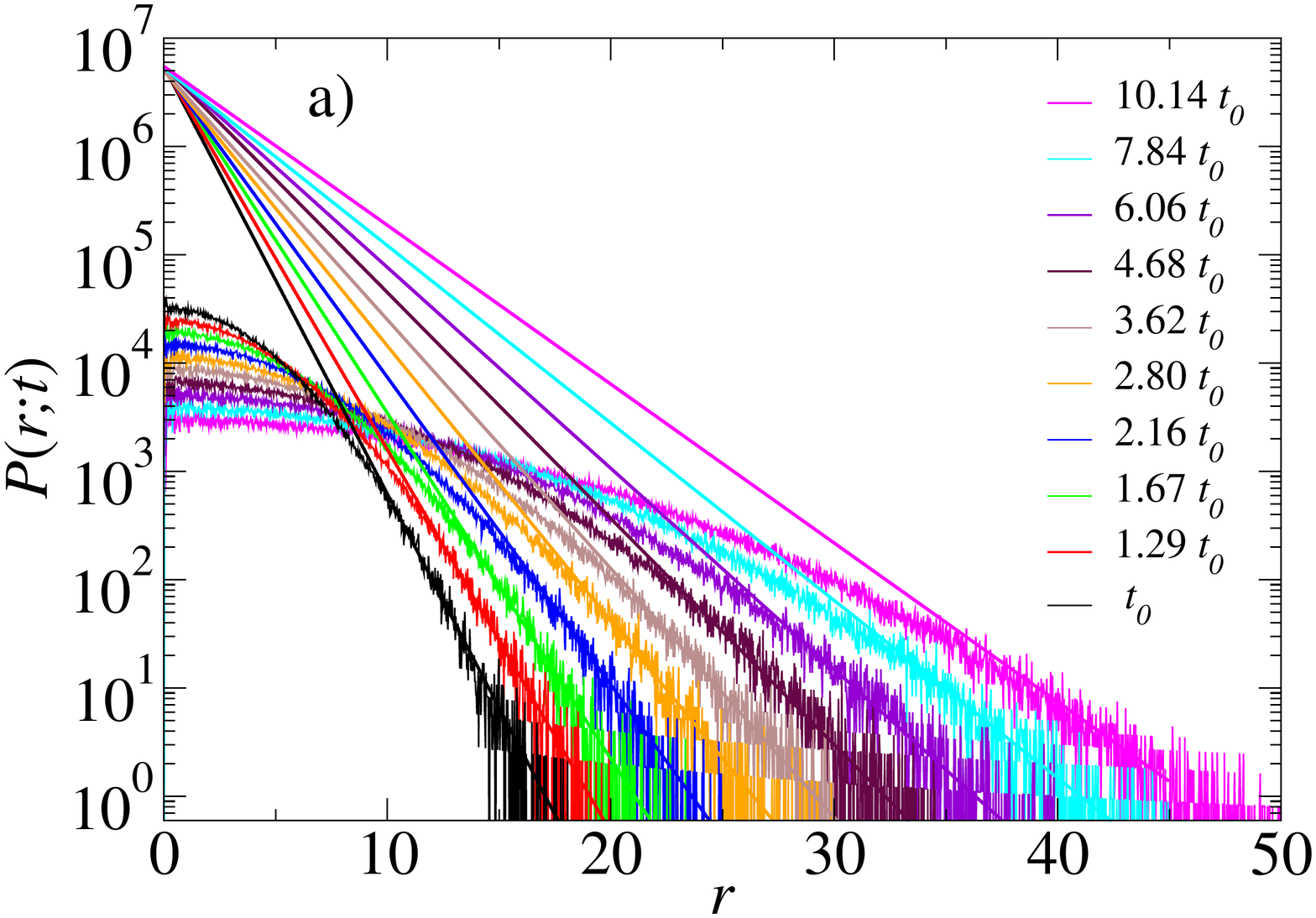}\\
\includegraphics[width=0.47\linewidth]{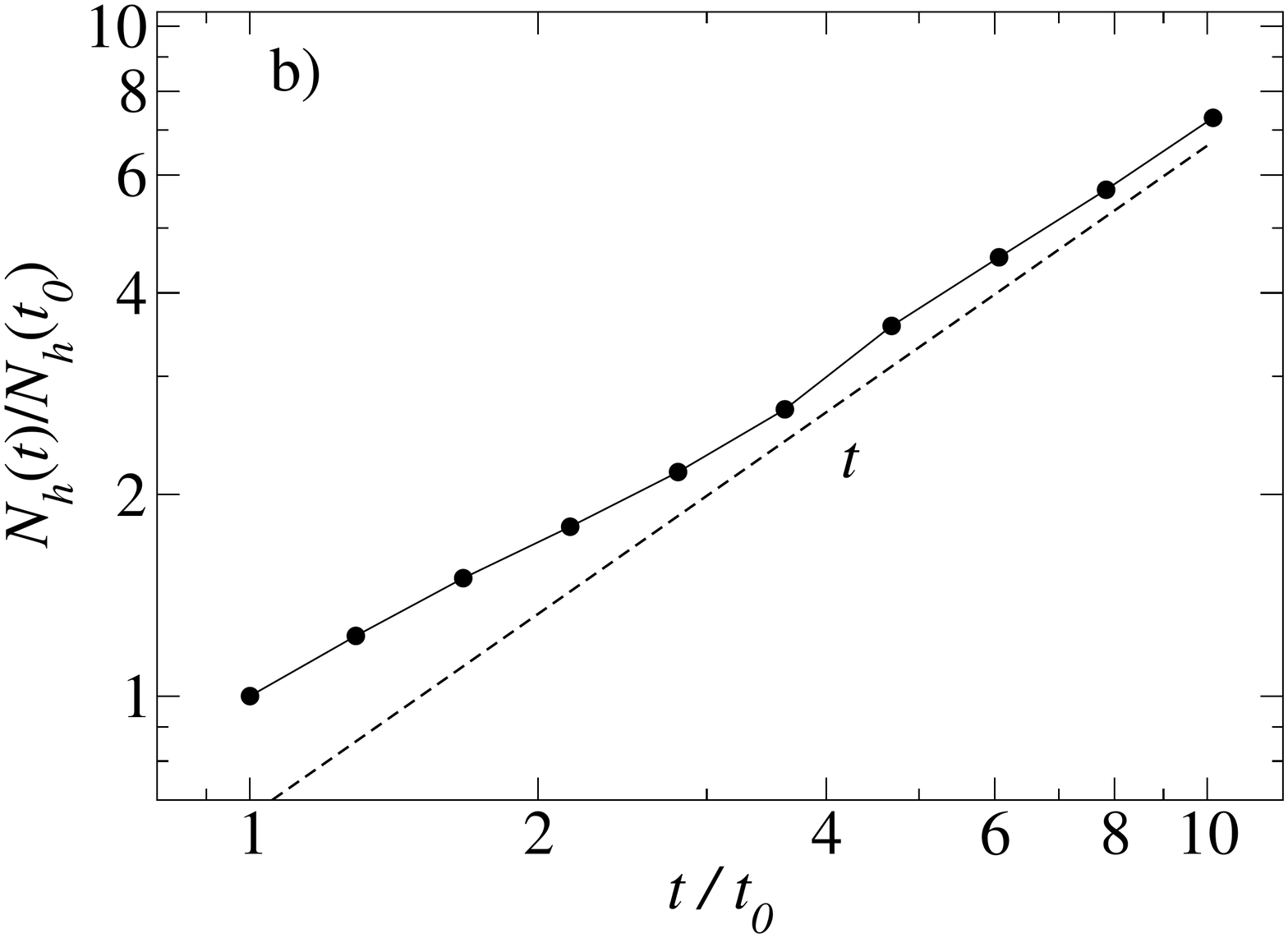}
\includegraphics[width=0.47\linewidth]{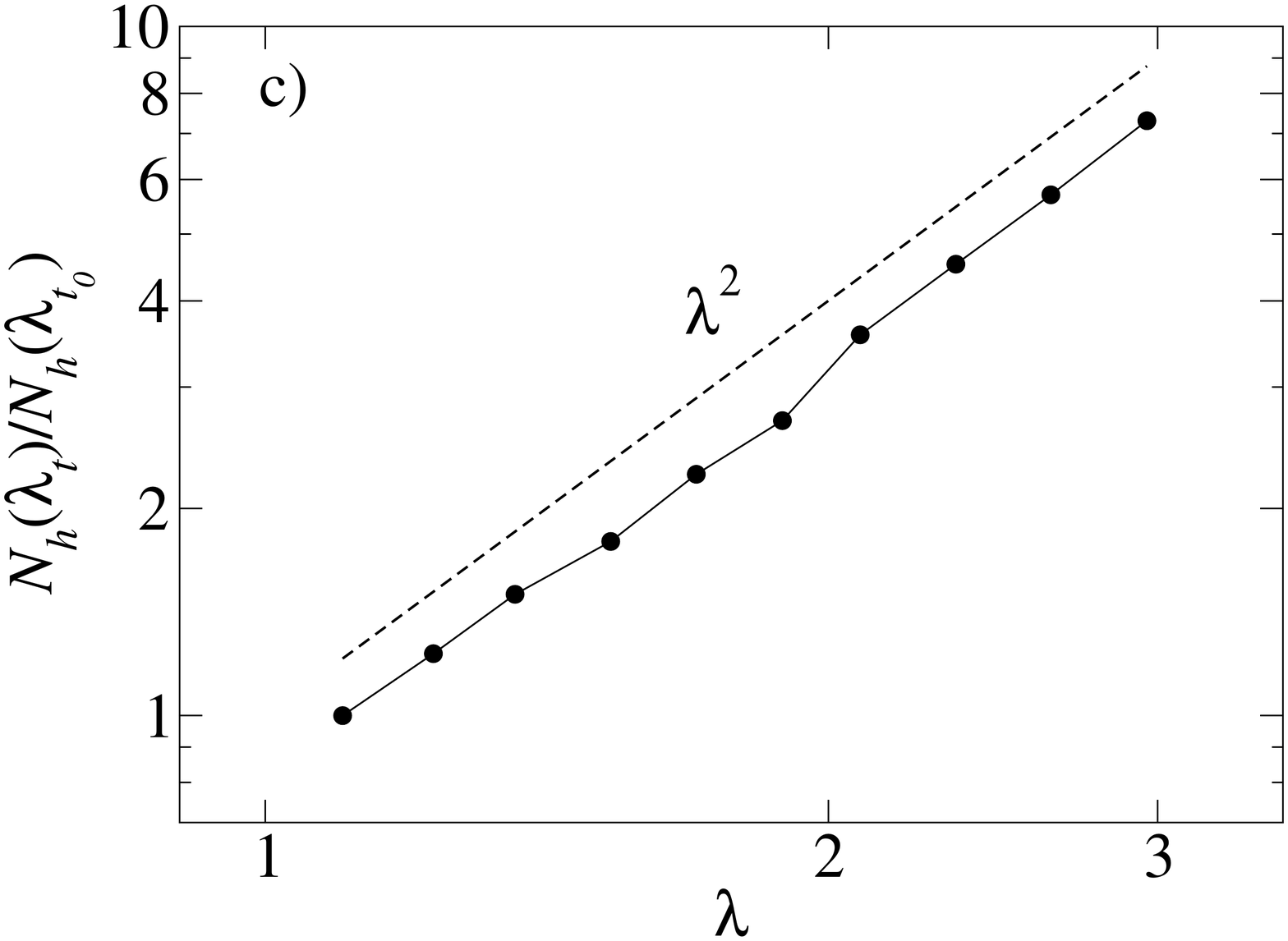}
\caption{Testing the scaling behavior for the 2D-Kob-Andersen binary mixture at $T = 0.36$: a) $P(r;t)$ at different times $t$ ($> t_0$); b) $N_h(t)$ normalized by $N_h(t_0)$, showing the expected linear behavior in time; c) $N_h(\lambda)$ normalized by $N_h(\lambda_{t_0})$, with the predicted $\lambda^2$ behavior.}
\end{figure*}

\end{document}